\documentclass[12pt]{iopart}

\usepackage[dcucite]{harvard}

\usepackage{graphicx}
\usepackage{times}

\renewcommand{\d}{\mathrm{d}}

\def\zsun{{\rm Z_\odot}}
\def\msun{{\rm M_\odot}}
\def\msunh{{\rm M_\odot/{\it h}}}
\def\fnl{$f_{\rm NL}$}
\def\Mpch{{Mpc/{\it h}}}

\begin{document}

\title[Non-Gaussianities and baryons]{
Gas distribution, metal enrichment, and baryon fraction in Gaussian and non-Gaussian universes
}

\author{Umberto~Maio$^{1}$}

\address{
${}^1$Max-Planck-Institut f\"ur extraterrestrische Physik, Giessenbachstra{\ss}e 1,  D-85748 Garching b. M\"unchen, Germany
}
\ead{umaio@mpe.mpg.de}

\begin{abstract}
We study the cosmological evolution of baryons in universes with and without primordial non-Gaussianities via N-body/hydrodynamical simulations, including gas cooling, star formation, stellar evolution, chemical enrichment from both population III and population II regimes, and feedback effects.
We find that large \fnl{} values for non-Gaussianities can alter the gas probability distribution functions, the metal pollution history, the halo baryon, gas and stellar fractions, mostly at early times.
More precisely:
(i) non-Gaussianities lead to an earlier evolution of primordial gas, structures, and star formation;
(ii) metal enrichment starts earlier (with respect to the Gaussian scenario) in non-Gaussian models with larger \fnl;
(iii) gas fractions within the haloes are not significantly affected by the different values of \fnl, with deviations of $\sim 1\%-10\%$;
(iv) the stellar fraction is quite sensitive to non-Gaussianities at early times, with discrepancies reaching up to a factor of $\sim 10$ at very high $z$, and rapidly converging at low $z$;
(v) the trends at low redshift are independent from \fnl{}: they are mostly led by the ongoing baryonic evolution and by the feedback mechanisms, which determine a  $\sim 25\%-30\%$ discrepancy in the baryon fraction of galaxy groups/clusters with respect to the cosmic values;
(vi) non-Gaussianity impacts on the cluster X-ray emission or on the SZ effect(s) are expected to be not very large and dominated by feedback mechanisms, whereas some effects on the 21-cm emission can be expected at early times;
(vii) in order to address non-Gaussianities in the cosmological structure contest, high-redshift ($z\sim 10$) investigations are required: first stars, galaxies, quasars, and GRBs may be potential cosmological probes of non-Gaussianities.
\end{abstract}





\section{Introduction}\label{Sect:introduction}
Any cosmological model has to face two main concerns: the first is to explain the space-time evolution of the Universe, and the second is to explain the formation and growth of the cosmic ``space-time content'', i.e. the cosmic structures.
\\
Nowadays, it is possible to give good constraints on the first problem \cite[and references therein]{wmap7_2010,Komatsu2011}: in particular, we know that the cosmological expansion is well described by the so-called $\Lambda$CDM model, in which the space-time is flat and its expansion rate at the present is H$_0=70\,\rm km/s/Mpc$ ($h=0.7$, if H$_0$ is normalized to $100~\,\rm km/s/Mpc$).
The density parameters describing the composition of the Universe are \cite{Komatsu2011} $\Omega_{\rm 0,m}=0.272$ for matter, and $\Omega_{\rm 0,\Lambda}=0.728$ for the cosmological constant $\Lambda$, often attributed to some kind of still unknown ``dark-energy'' \cite[firstly investigated its effects on gas hydrodynamics and cosmic chemical evolution via detailed numerical simulations and found some impacts on the early gas cloud distributions]{Maio2006}.
According to the cited observational data, the resulting total-density parameter is unity, $\Omega_{\rm 0,tot}=1.0$, but the only physically known fraction is very small, since baryons yields only $\Omega_{\rm 0,b}=0.0456$.
\\
The second problem is fundamentally related to the early ``seeds'' from which cosmological structures will develop.
More precisely, clusters and galaxies are supposed to evolve from primordial matter fluctuations, that have a power-law primordial power spectrum, normalized via mass variance within $\rm 8\,\Mpch$-radius sphere $\sigma_8=0.8$, and with a spectral index $n=0.96$.
Then, in order to make statistical predictions for the structure growth, such as mass functions or bias \cite[for example]{Wagner2010,DAmico2011,Cosimo2011,Hamaus2011arXiv}, one has to assume a probability distribution of primordial fluctuations, around the mean density $\langle\rho\rangle$, as a function of the overdensity $\delta$.
According to the central-limit theorem, $\delta$ is usually assumed to be Gaussian.
However, the validity of this assumption has often been questioned over the years and non-Gaussian models have been considered since long time \cite{Peebles1983,GrinsteinWise1986}.
Recent observational data do not rule out non-Gaussianities either \cite{Komatsu2005,Komatsu2010}.
\\
The most common way to parameterize non-Gaussian distributions \cite[and references therein]{DesjacquesSeljak2010,Verde2010} is by introducing a {\it non-linearity} parameter, \fnl, according to the following {\it Ansatz} \cite{Salopek1990}:
\begin{equation}
\label{eq:Def}
\Phi = \Phi_{\rm L} + f_{\rm NL} \left[\Phi_{\rm L}^2 -\langle \Phi_{\rm L}^2\rangle \right].
\end{equation}
In the previous equation (\ref{eq:Def}), $\Phi$ is the Bardeen gauge-invariant potential, $\Phi_{\rm L}$ is the ``linear'' Gaussian part and the second term on the right-hand side is the non-linear correction, whose contribution is regulated by the \fnl{} parameter.
Since at any point the potential $\Phi$ depends only on the value of $\Phi_{\rm L}$, non-Gaussianities of this form are also called ``local''.
Observational constrains on primordial matter fluctuations suggest that the primordial distribution could present some departures from the Gaussian one, with \fnl{} in the range $\sim 10-100$ \cite[and their Table~2 for a collection of constraints]{MaioIannuzzi2011}.
The existence of a non-null \fnl{} would reflect on the growth of cosmological structures at any time and redshift ($z$).
This means variations of the dark-matter halo mass function, mostly at large masses and/or at high redshift \cite[and references therein]{DesjacquesSeljak2010,Verde2010,LoVerde2011arXiv,Desjacques2011arXiv}.
Non-Gaussian effects could also be visible in the baryon history of cosmic objects \cite{MaioIannuzzi2011} and probably affect their gaseous and stellar components \cite{MaioKhochfar2011}.
The uncertainties on the baryonic evolution in non-Gaussian scenarios are mainly linked to the feedback effects which could strongly interfere with the original non-Gaussian signal.
Besides the first attempt by \cite{MaioIannuzzi2011}, there are, at the moment, no other investigations in this field, and quite a large lack of knowledge of the hydrodynamical interplay of the cosmic gas with non-Gaussianities.
The ignorance of the detailed gas properties and of feedback effects could be a strong limitation \cite{et2011} when accounting for, e.g., metal abundances, Sunyaev-Zeldovic (SZ) signal \cite{Sunyaev1969,Sunyaev1970}, or X-ray emission from galaxy clusters in non-Gaussian models.
Additionally, it might also influence the gas emission at high redshift (like the 21-cm signal).
\\
Therefore, in the present study we analyze the principal features of the gas behaviour, during its condensation process within dark-matter haloes and address the effect on the baryon fraction in the Gaussian and in different non-Gaussian scenarios.
We will focus on the cosmic gas and its distribution at high and low redshift, down to $z=0$, by stressing the consequences from star formation, and the impacts for metal spreading, gas fractions, stellar fractions and baryon evolution within the formed cosmological structures.
\\
This work is organized as follows: in Sect.~\ref{Sect:simulations}, we summarize the main properties of the simulations and of the physical implementations; in Sect.~\ref{Sect:results}, we show our results about star formation (Sect.~\ref{sect:sfr}), gas probability distribution functions (Sect.~\ref{sect:pdf}), metal enrichment (Sect.~\ref{sect:ff}), and halo baryonic properties (Sect.~\ref{sect:baryons}); we finally discuss our results and summarize our conclusions in Sect.~\ref{sect:discussion}.


\section{Simulations}\label{Sect:simulations}

We use the simulations presented by \cite{MaioIannuzzi2011}, who performed the first study of the effects of primordial non-Gaussianities on the cosmic star formation, via self-consistent N-body hydrodynamical chemistry simulations.
Beyond gravity and hydrodynamics, the code includes radiative gas cooling at high \cite{SD1993} and low \cite{Maio2007} temperatures, a multi-phase model for star formation, UV background radiation, and wind feedback \cite[for further details]{Katz1996,Springel2003}.
It also follows chemical evolution, metal pollution from population III (popIII) and/or population II (popII) stellar generations, ruled by a critical metallicity threshold of $Z_{crit}=10^{-4}\,\zsun$ \cite{Tornatore2007,Maio2009,Maio2009PhDT,Maio2010,Maio2011b}.
The main agents of gas cooling \cite[e.g.]{SD1993,Maio2007} are atomic resonant transitions of H and He excitations that can rapidly bring temperatures down to $\sim 10^4\,\rm K$, when molecules become the only relevant coolants in pristine environments.
In metal-rich environments, atomic fine-structure transitions are additionally included \cite{Maio2007} and can further lower them, mostly at high redshift, since the formation of a UV background \cite{HaardtMadau1996,Dave1999,Gilmore2009} at lower redshifts heats the IGM and hinders gas cooling.
In dense environments, stochastic star formation is assumed, cold gas is gradually converted into stars \cite{Kennicutt1998}, outflows take place, and feedback effects inject entropy into the surrounding environment, leading to a self-regulated star forming regime \cite[for further detail]{Springel2003}.
\\
The simulations considered here assume a concordance $\Lambda$CDM model with
total density parameter  $\Omega_{\rm 0,tot}=1.0$,
matter density parameter $\Omega_{\rm 0,m}=0.3$,
cosmological density parameter  $\Omega_{\rm 0,\Lambda}=0.7$,
baryon density parameter $\Omega_{\rm 0,b}=0.04$,
expansion rate at the present H$_0=70\,\rm km/s/Mpc$,
power spectrum normalization via mass variance within $\rm 8\,\Mpch$-radius sphere $\sigma_8=0.9$,
and spectral index $n=1$.
We consider the cases: \fnl = 0 (Gaussian), 100, 1000, in cosmological periodic boxes with 100~\Mpch{} a side, and with a gas mass resolution of $\sim 3\times 10^8\,\msunh$.
\\
The stellar initial mass function (IMF) is a key ingredient for stellar systems, because it determines the temporal evolution and the relative distribution of stars with different masses and different metal yields.
In the simplest approximation, it can be parameterized as a power law of the mass converted into stars $M_\star$ \cite{Salpeter1955}
\begin{equation}
\label{eq:imf}
\phi(M_{\star})\propto M_{\star}^x
\end{equation}
and normalized via the relation
\begin{equation}
\label{eq:imf}
\int_{M_{\star,inf}}^{M_{\star,up}} \phi(M_{\star}){\rm d} M_\star = 1
\end{equation}
over the mass range [$M_{\star,inf}, M_{\star,up}$].
Since the primordial IMF is quite uncertain, we bracket the possible scenarios by assuming both a top-heavy popIII IMF (normalized over the mass range [100, 500] solar masses, and with metal yields for PISN), and a \cite{Salpeter1955}-like IMF (normalized over the mass range [0.1, 100] solar masses, and with metal yields for AGB, SNII, SNIa stars), as opposite limiting cases \cite{MaioIannuzzi2011}.
The slope is fixed to $x=-2.35$ \cite{Salpeter1955}.
For the popII-I stellar populations, we always adopt a common Salpeter IMF.


\section{Results}\label{Sect:results}
In the next sections we show the main results of our analysis, starting, for clarity, from the star formation rate densities in the simulated volumes (Sect.~\ref{sect:sfr}) and considering the implications for the basic baryon properties, like probability distribution functions (Sect.~\ref{sect:pdf}), and metal enrichment (Sect.~\ref{sect:ff}).
Then we will analyze the consequences on the cosmological structures (Sect.~\ref{sect:baryons}), exploring the evolution of their gaseous and stellar components (Sect.~\ref{sect:evolution}, \ref{sect:massivehalo}, and \ref{sect:profiles}).

\subsection{Star formation}\label{sect:sfr}

\begin{figure*}
\centering
\includegraphics[width=0.495\textwidth]{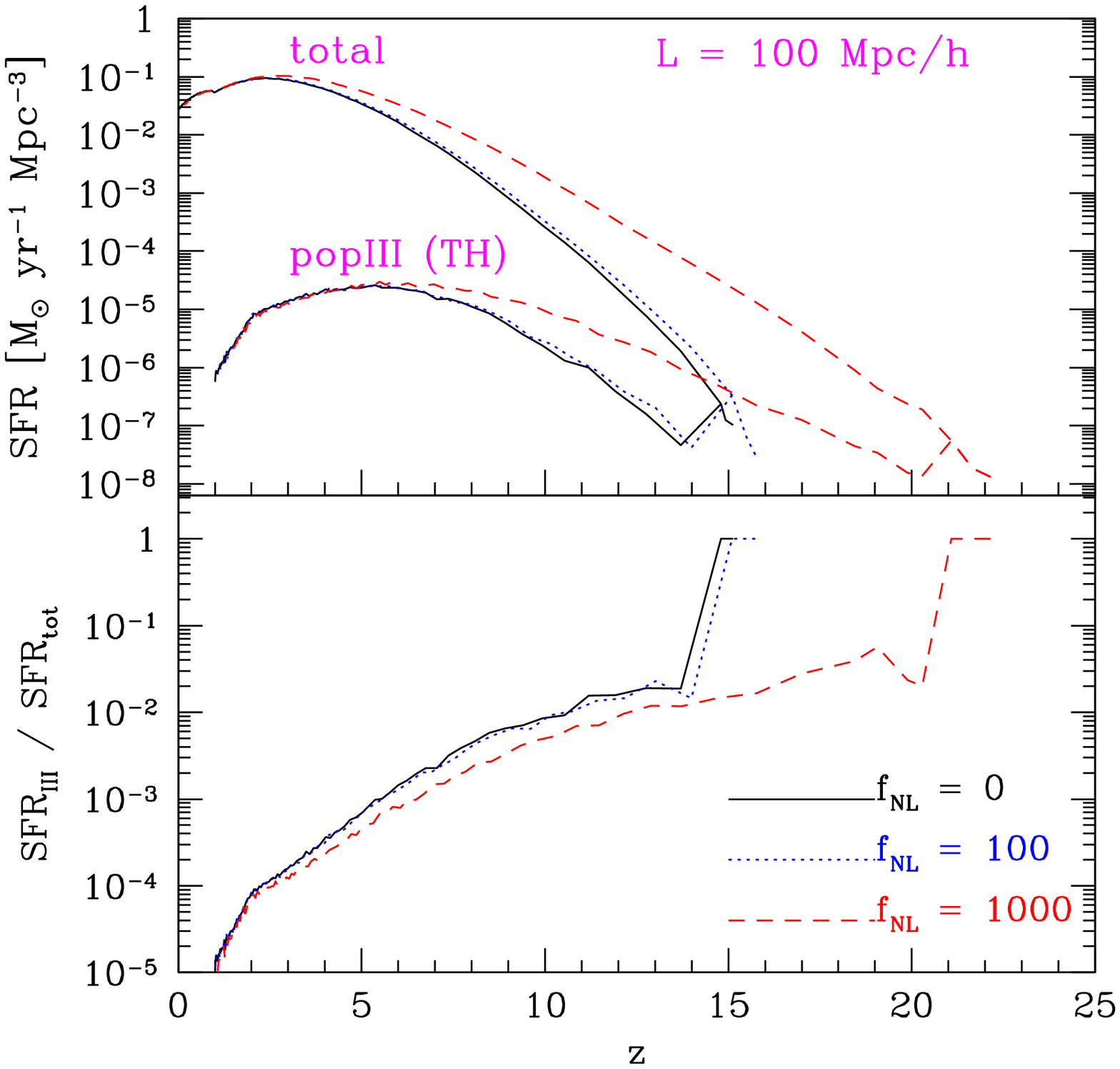}
\includegraphics[width=0.495\textwidth]{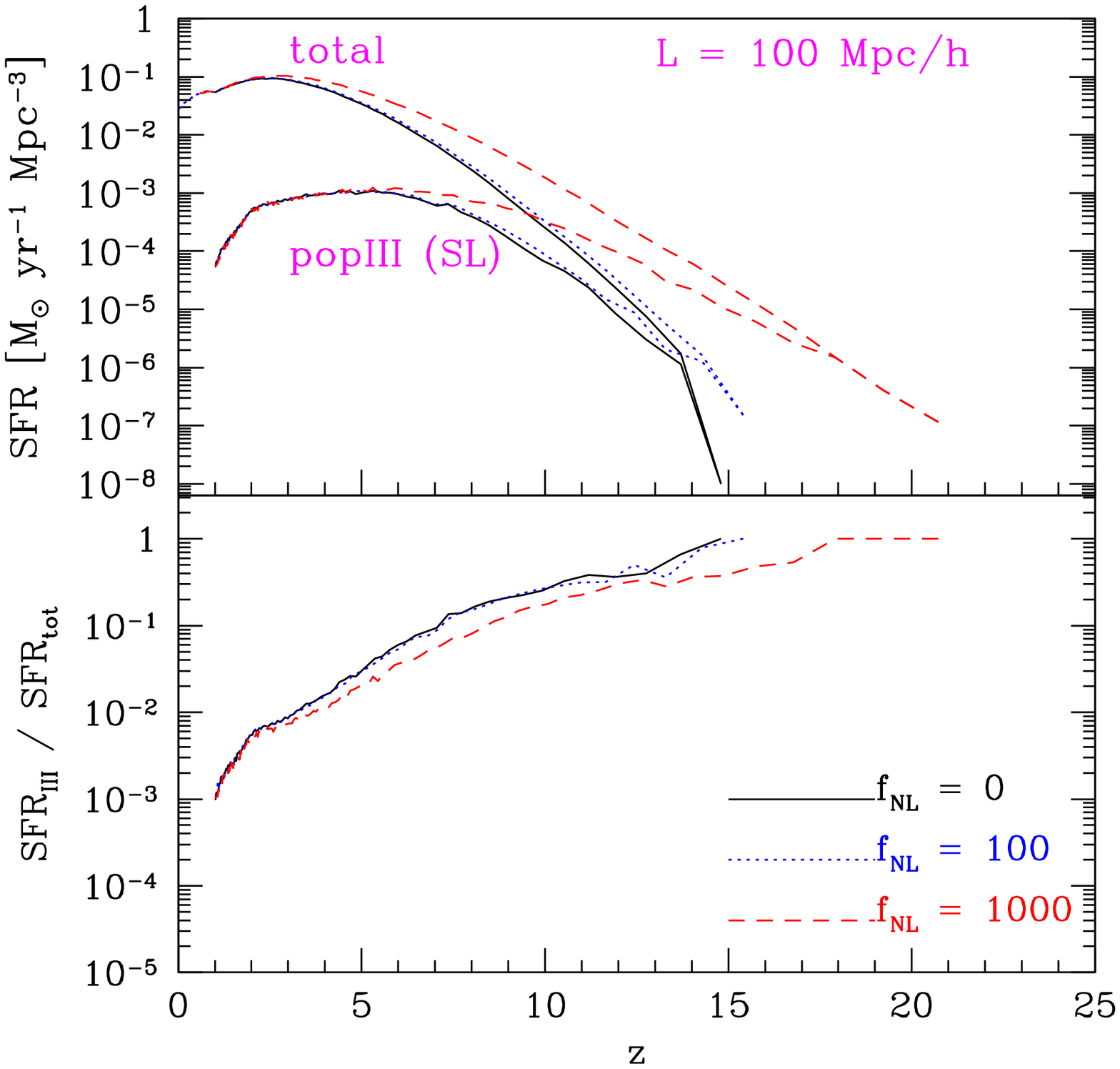}
\caption[SFR evolution]{\small
Star formation rate densities as function of redshift for \fnl=0 (solid lines), \fnl=100 (dotted lines), and \fnl=1000 (dashed lines).
Individual contributions from popIII regions (bottom lines) are also shown.
In the left case the assumed popIII IMF is top-heavy (TH) in the mass range [100,500]~$\rm M_\odot$, with a slope of -1.35 \cite[have showed that the detailed value of the slope has little effects]{Maio2010}, while in the right case the assumed popIII IMF is Salpeter like (SL) \cite[for further details]{MaioIannuzzi2011}.
}
\label{fig:SFR}
\end{figure*}
We already mentioned that the N-body/hydrodynamical/chemistry simulations considered here have been presented for the first time in \cite{MaioIannuzzi2011}, who discussed the main effects of non-Gaussianities on the star formation history in the Universe.
In the present work we aim at a more detailed study of the cosmological evolution of cosmic structures, and in particular of their visible components, gas and stars, within Gaussian and non-Gaussian contexts.
Therefore, we start from the star formation rate densities,which are re-proposed, for sake of completeness, in Fig.~\ref{fig:SFR}, for both the top-heavy popIII IMF (left) and the Salpeter-like popIII IMF (right).
The star formation rate density, $\dot\rho_{\star}$, quantifies the mass converted into stars, $M_\star$, per unit time, $t$, and unit volume, $V$, and can be aspressed as \cite{Kennicutt1998}
\begin{equation}
\label{eq:sfr}
\frac{{\rm d}\rho_{\star}}{{\rm d}t} = \frac{{\rm d} M_\star}{{\rm d} t~{\rm d} V} \propto \frac{\rho}{t_{ff}},
\end{equation}
where $\rho$ is the gas density and $t_{ff}\sim 1/\sqrt{G\rho}$ is the free-fall time.
Therefore,the larger the gas density the larger the star formation rate density.
As clear from the plots in Fig.~\ref{fig:SFR}, the dependences on \fnl{} are quite weak and it is not easy to distinguish among different cases, as long as \fnl$\le 100$.
Only an extremely high (and currently ruled-out) value of \fnl=1000 would show appreciable differences with respect to the Gaussian case (\fnl=0).
\\
Moreover, the global trends show that the Gaussian models give a fair representation of the star formation history, also with respect to \fnl=100 non-Gaussian corrections.
The main difference is found for the \fnl=1000 case, which predicts a significantly earlier onset of star formation of almost $10^8\rm yr$, and, thus, a correspondingly earlier enrichment process from the first stars and galaxies (see following sections).
\\
The adopted primordial popIII IMF has consequences on the popIII star formation.
It determines a popIII contribution to the total star formation rate varying between almost two orders of magnitudes.
More in detail, in the top-heavy popIII IMF case, the popIII star formation rate densities reach at most $\sim 10^{-5}\,\rm\msun/yr/Mpc^3$, for any \fnl{}, and the relative contributions drop down below $\sim 10^{-5}$ at low redshift.
In the Salpeter-like popIII IMF case, instead, the star formation rate densities achieved are $\sim 10^{-3}\,\rm\msun/yr/Mpc^3$, with a popIII contribution of less than $\sim 10^{-3}$, even at low redshift, independently from \fnl.
\\
We stress that the behaviour of the total star formation rate density at low $z$ is practically unaffected by deviations from Gaussianity.

\subsection{Gas distribution functions}\label{sect:pdf}

\begin{figure*}
\centering
\includegraphics[width=0.32\textwidth]{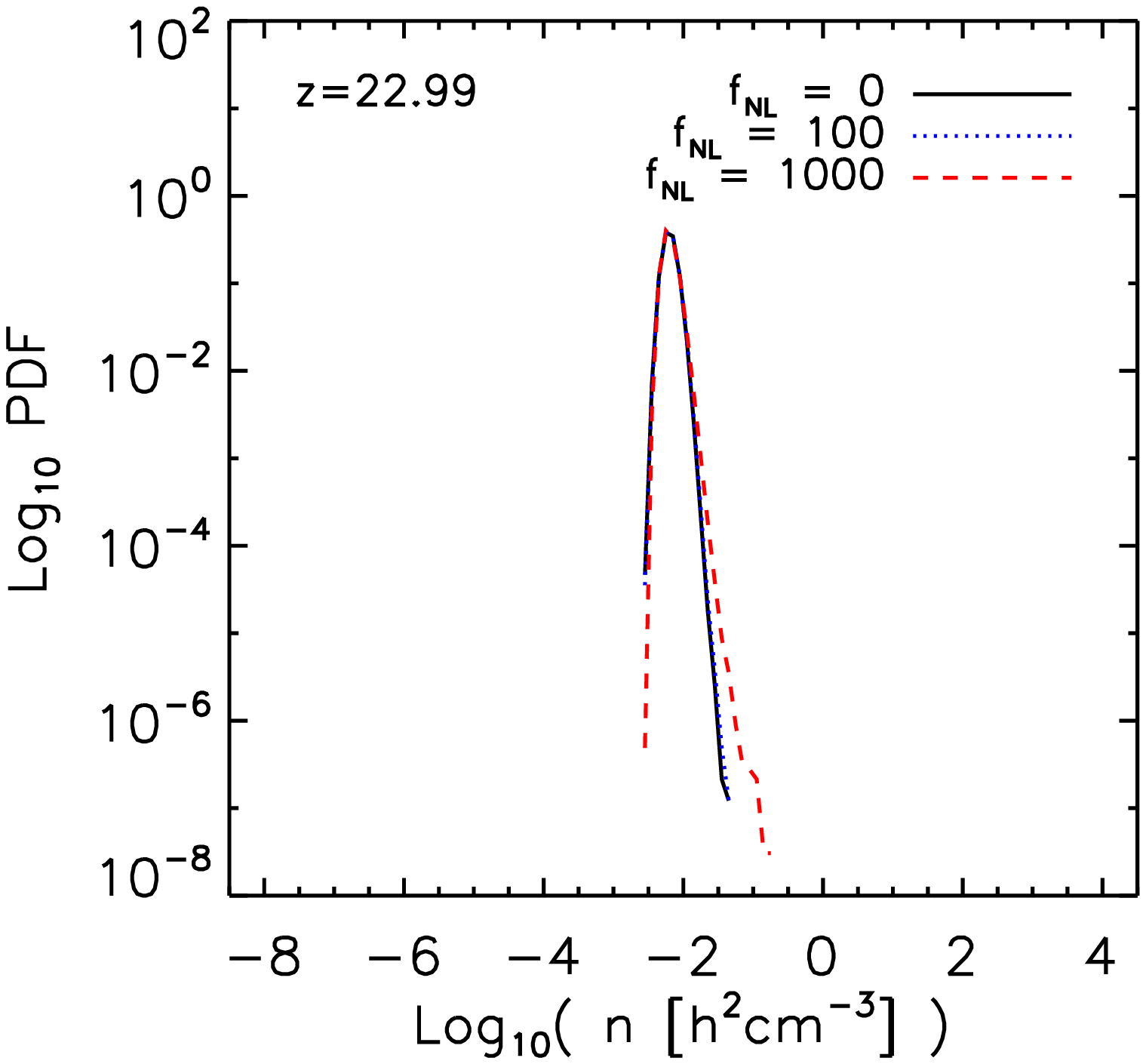}
\includegraphics[width=0.32\textwidth]{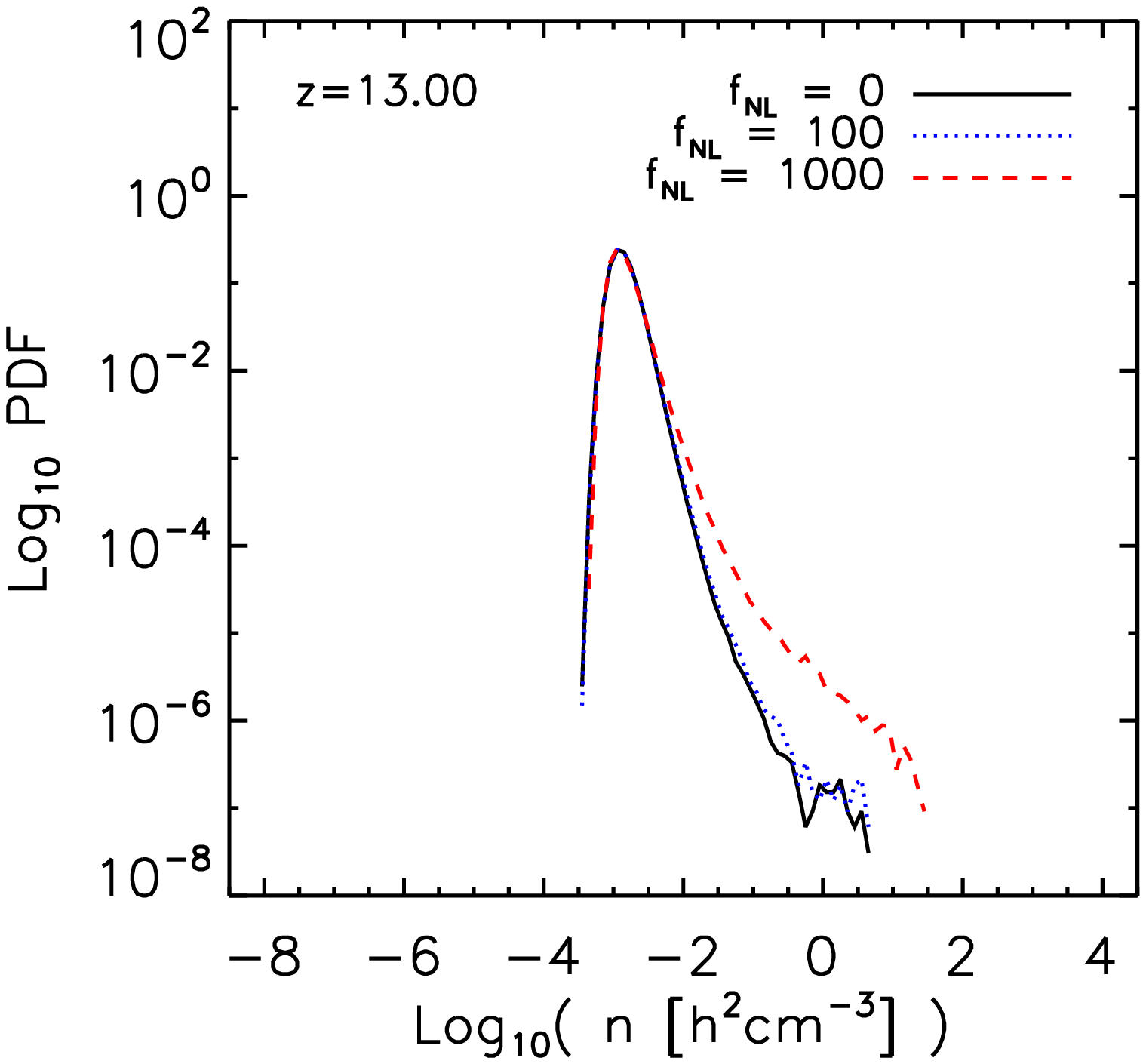}
\includegraphics[width=0.32\textwidth]{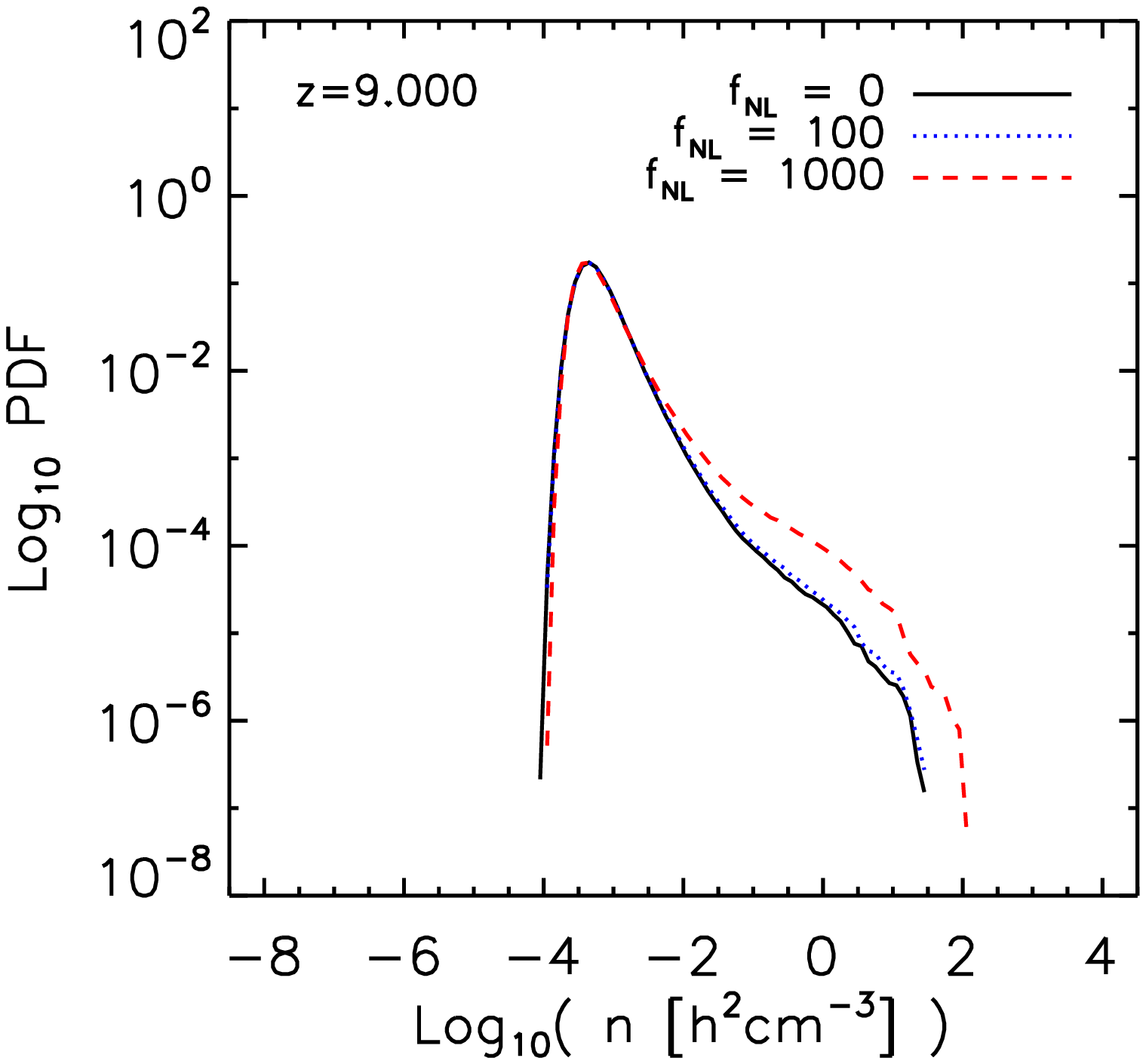}\\
\includegraphics[width=0.32\textwidth]{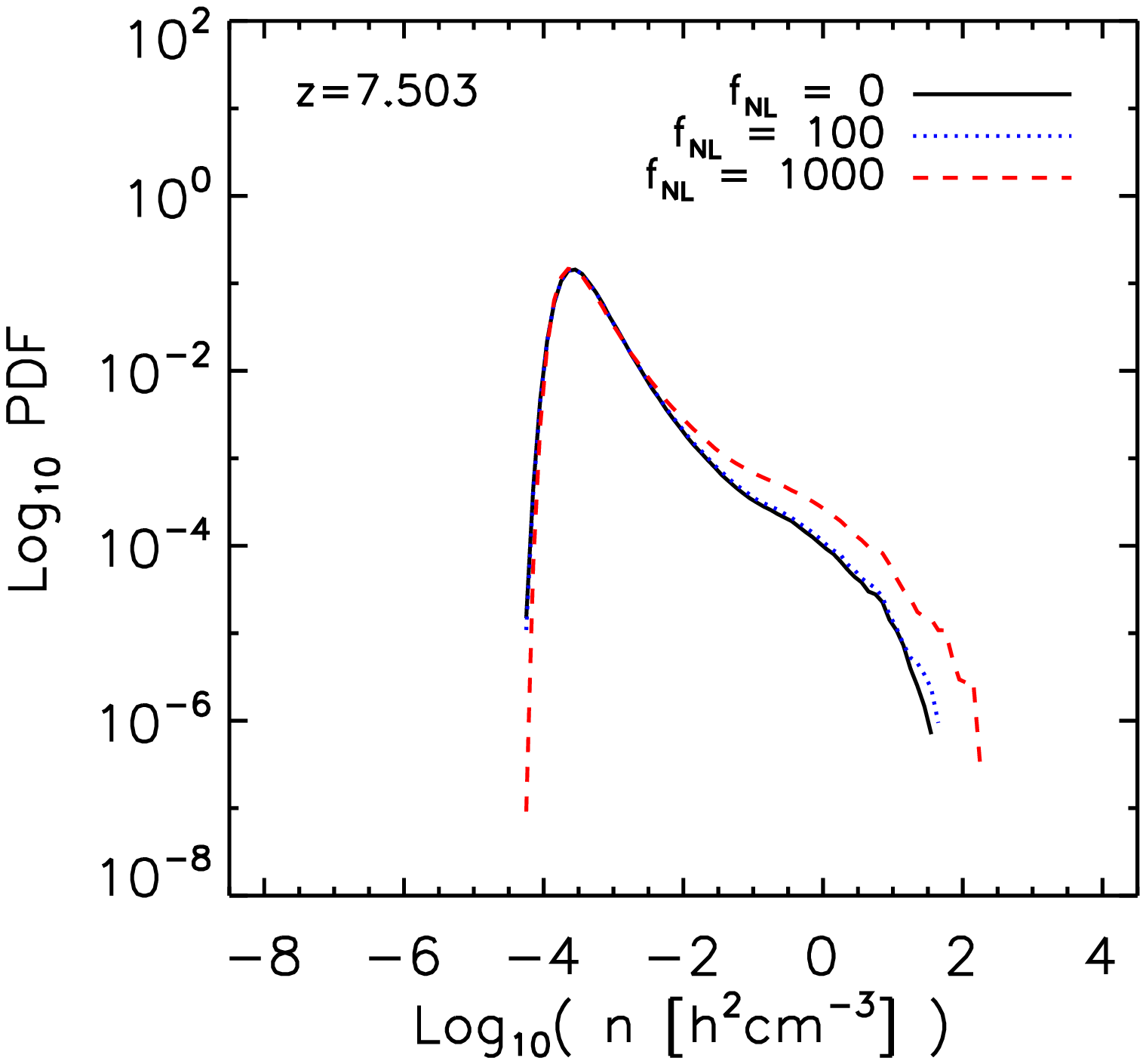}
\includegraphics[width=0.32\textwidth]{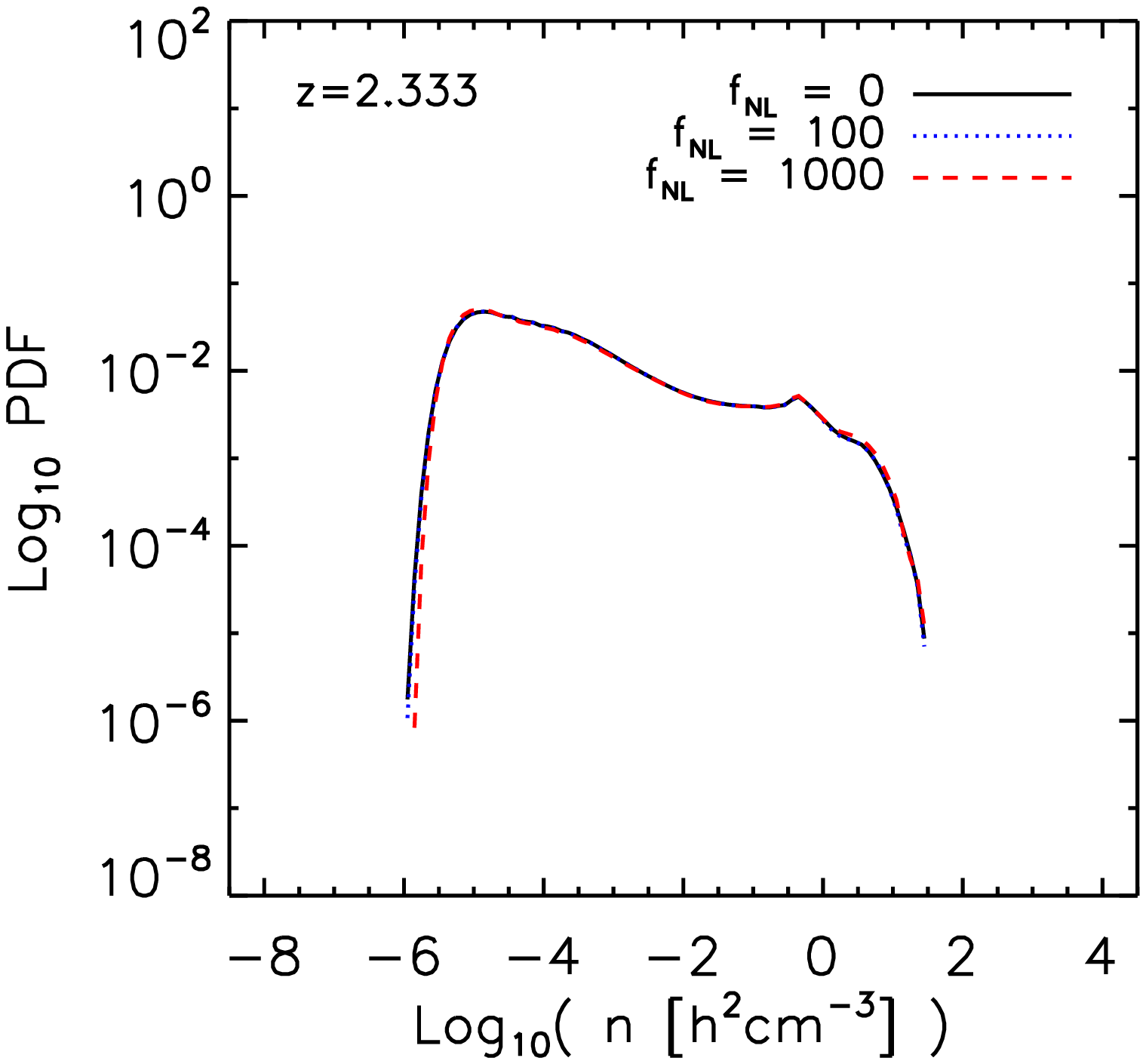}
\includegraphics[width=0.32\textwidth]{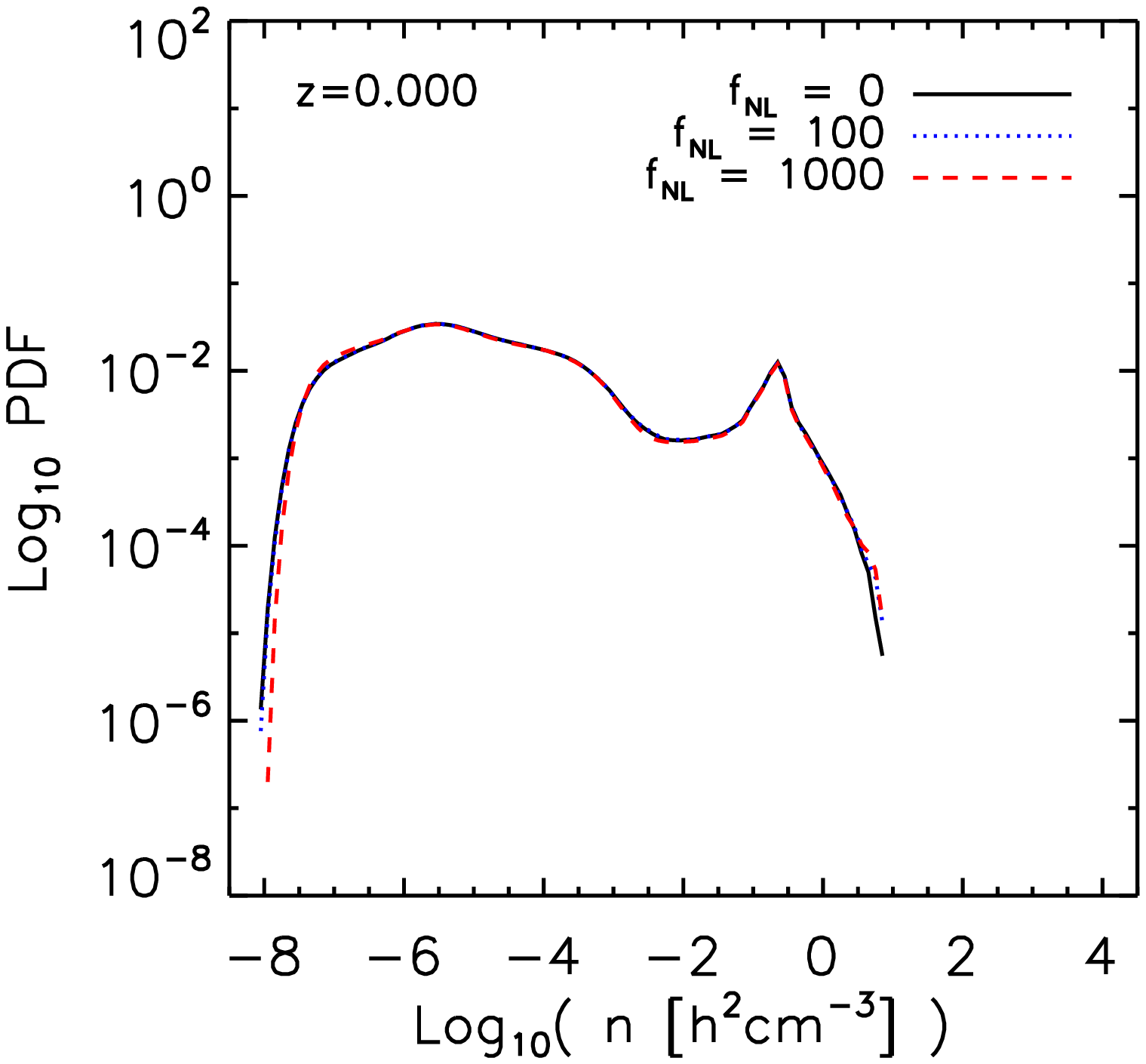}
\caption[Redshift comparison]{\small
Gas probability distribution function at different redshift, $z$, as indicated by the labels, for the boxes with $100\,\rm \Mpch$ a side, and a top-heavy popIII IMF.
}
\label{fig:pdf100}
\end{figure*}

Star formation is the main agent leading the whole evolution of cosmic gas.
Indeed, at very early times, when the Universe is younger than a few $10^8\,\rm yr$, there are no relevant star formation episodes, yet, and the gas has a purely pristine chemical composition.
First stars synthesize heavy elements in their cores and eventually spread them around, during the final stages of their lives, by exploding as SN or PISN.
These explosive events lead metal pollution in the surrounding regions and inject large amounts of entropy into the medium.
Obviously, this has severe impacts (feedback) on the following generations, since the gas will be in very different conditions, with respect to the primordial, pristine gas.
In order to illustrate and summarize the behaviour of cosmic gas during the cosmological evolution and its modifications due to non-Gaussianities, we consider the probability distributions, $P(n)$, as functions of the number density, $n$:
\begin{equation}
\label{eq:pdf_diff}
P(n) = \frac{{\rm d} f}{{\rm d} n}
\end{equation}
with $\d f$ fraction of gas elements in the $\d n$ density bin, and with normalization
\begin{equation}
\label{eq:pdf_integral}
\int_{-\infty}^{+\infty} P(n){\rm d} n = 1.
\end{equation}
We display the gas probability distribution functions in Fig.~\ref{fig:pdf100}, at different redshifts, for the runs having \fnl=0, \fnl=100, and \fnl=1000. The assumed popIII IMF is top-heavy.
From the plots, it is evident that at early times (e.g. at $z\sim 23-13$) the gas distributions are slightly affected by non-Gaussianities, the gas is mainly at the mean density, and the environments showing some deviations from the Gaussian behaviour are only the ones with over-densities $\sim 10^1-10^2$, for \fnl=1000.
This basically reflects the features of the initial conditions, that, in the non-Gaussian cases, are biased towards higher densities.
Within $\sim 1\,\rm Gyr$, a decrement in the differences of the high-density tails appears, with discrepancies going from $\sim 1$ order of magnitude at $z\sim 13$ down to a factor of $\sim 2-3$ at $z\sim 7.5$.
Moreover, below $z\sim 7.5$, the ongoing star formation events and the related feedback effects efficiently push the material far away from the star forming sites: this causes a larger spread in the distributions of the underdense material that is fed by gas expelled by the dense star forming regions.
The feedback mechanisms cause gas mixing among the cold and hot phases, and the interactions between the inflowing and outflowing material increase the turbulent state up to Reynolds number of the order of $\sim 10^8-10^{10}$ \cite{Maio2011b}.
These phenomena completely erase any imprints of non-Gaussianities (already by $z\sim 3$), for any \fnl.
This is an important conclusion, since it clearly shows that low-$z$ baryon evolution is affected by gas and stellar feedback in a much more important way than by deviations from the Gaussian shape of the underlying matter distribution (see further discussion in Sect.~\ref{sect:massivehalo}), even when considering very large \fnl.
Gas emission at high redshift (e.g. the Hydrogen hyperfine-structure transition at 21~cm from early structures) is not supposed to change strongly for \fnl=0 and \fnl=100.
For \fnl=1000, whereas, there would be an enhancement due to the slightly larger amounts of gas at higher densities, but only within the first Gyr.
\\
The Salpeter-like popIII IMF cases present similar physical behaviours, with distributions which change accordingly to the different \fnl{} values.
\\
More differences will be found in the metallicity evolution, as we will discuss in the next section.
\\
We note though (see also discussion in Sect.~\ref{sect:discussion}) the differences at early times in both gas distributions and star formation rates already suggests that primordial ``visible'' structures like first stars, galaxies, quasars, and gamma-ray bursts (GRBs) could be affected by the initial (Gaussian or non-Gaussian) dark-matter distribution.

\subsection{Metals}\label{sect:ff}

\begin{figure*}
\centering
\includegraphics[width=0.495\textwidth]{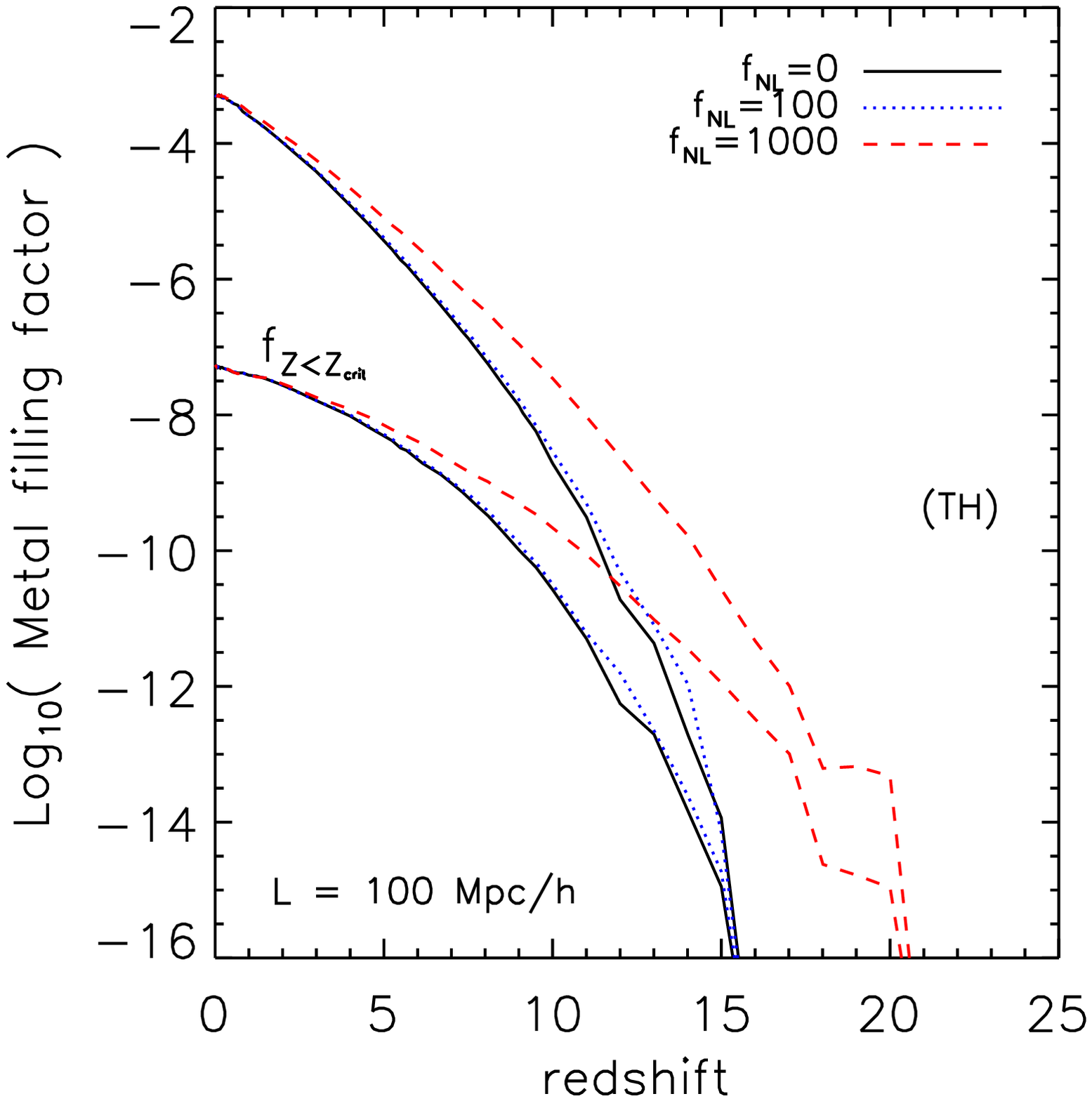}
\includegraphics[width=0.495\textwidth]{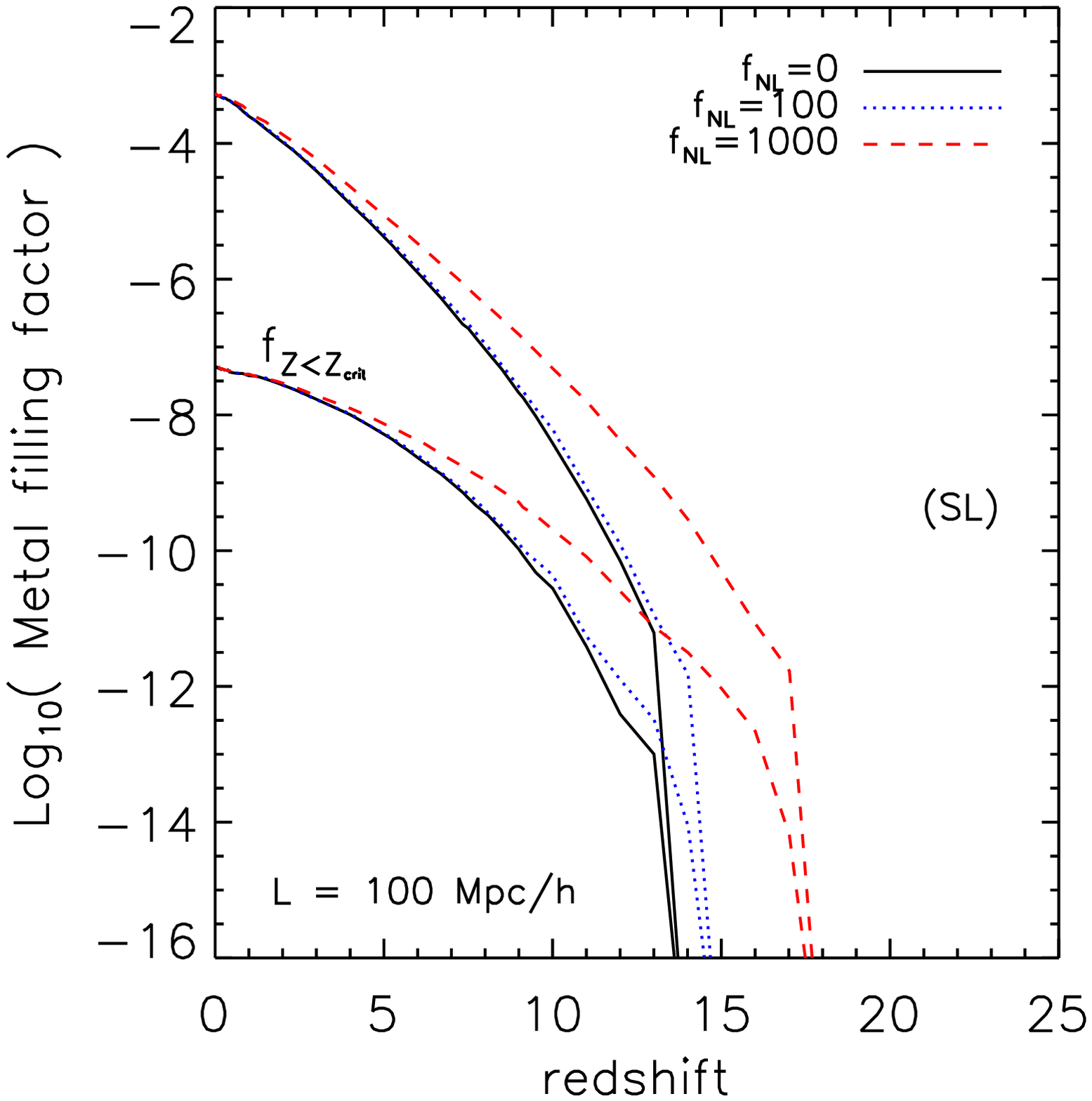}
\caption[Metal filling factor evolution]{\small
Metal filling factors as function of redshift in the 100~\Mpch{} side boxes for \fnl=0 (solid lines), \fnl=100 (dotted lines), and \fnl=1000 (dashed lines). Individual contributions from popIII regions ($0<Z<Z_{crit}$, bottom lines) are also shown.
In the left panel the assumed popIII IMF is top-heavy (TH) in the mass range [100,500]~$\rm M_\odot$, with a slope of -1.35 \cite[have showed that the detailed value of the slope has little effects]{Maio2010}, while in the right panel the assumed popIII IMF is Salpeter-like (SL).
}
\label{fig:ffIMF}
\end{figure*}

Metal spreading is one of the principal signatures of the evolution of baryonic matter, and it leads crucial changes to the formation process of cosmic objects.
\\
To show the effects of different \fnl{} values on metal evolution, in Fig.~\ref{fig:ffIMF} we present the metal filling factor computed for the runs with \fnl=0, \fnl=100, and \fnl=1000.
This quantity represents the fraction of cosmic volume filled by metals and is computed from the outputs of our simulations, at each redshift, according to the equation:
\begin{equation}
\label{eq:ff}
f_V \equiv \frac{\sum_i m_{Z,i}/\rho_{Z,i}}{\sum_j m_j/ \rho_j} \sim \frac{\sum_i m_{Z,i}/\rho_{Z,i}}{V},
\end{equation}
with $i$ and $j$ integers running over the number of the enriched particles and of all the SPH particles, respectively,
$m_{Z}$ particle metal mass,
$\rho_{Z}$ metal-mass density,
$m$ total particle mass,
$\rho$ total-mass density,
and $V$ simulation box volume.
On the left panel, we show results from the simulations with a top-heavy popIII IMF, while, on the right panel, we show results from the simulations with a Salpeter-like popIII IMF.
In the plots, we also distinguish by metallicity regime to study the effects on the different stellar populations.
The most evident feature is the convergence at low redshift of all the models.
The most relevant differences are at high redshift, in correspondence of the different onsets of the first star forming events.
\\
In the top-heavy popIII IMF case (left panel), the first enrichment patterns are found immediately after the first star formation events, due to the extremely short lifetimes of massive stars and PISN ($\sim 2\times 10^6\,\rm yr$).
Indeed, in the runs with \fnl=0 and \fnl=100 the onset is at $z\sim 16$, and in the run with \fnl=1000 it is at $z\sim 22$ (see left panel of Fig.~\ref{fig:SFR}).
Correspondingly, first metals are spread in the Universe at $z\sim 15$ for \fnl=0 and \fnl=100, and at $z\sim 21$ for \fnl=1000.
Given the large metal yields \cite[for further discussions]{Maio2010}, the popIII filling factors become soon sub-dominant of some orders of magnitudes, for all the cases.
We stress that the differences between the Gaussian case (\fnl=0) and the \fnl=100 case are negligible at any time.
At lower redshift, when effects of non-Gaussianities are less relevant, all the models predict a total volume metal filling factor of $\sim 10^{-3}$, and a popIII volume filling factor of $\sim 10^{-7}$, four orders of magnitude smaller.
\\
When a Salpeter-like popIII IMF is adopted (right panel), the main changes are found at high redshift and are mainly due to the longer stellar lifetimes.
In fact, they induce a larger delay of the pollution episodes, since the first metals are spread at $z\sim 14$ for \fnl=0 and \fnl=100, and at $z\sim 18$ for \fnl=1000, i.e., about $\sim 5\times 10^7\,\rm yr$, after the onset of star formation (consistently with the typical lifetimes of the first SNII explosions).
Also in these scenarios differences between \fnl=0 and \fnl=100 are not very appreciable and the asymptotic trends, led mostly by the popII-I regime, remain unaltered.

\subsection{Baryon fraction in the cosmic haloes}\label{sect:baryons}
After the general, global quantities previously studied, we investigate if and how the situation changes in the individual haloes found in the simulations, by paying more attention to their gaseous and stellar content.
The formed structures are identified by a friends-of-friends (FoF) algorithm, with a linking length of $20\%$ the mean inter-particle separation.
In this way, we can determine total masses $M_{tot}$ and basic properties of all the cosmic objects and of their different dark-matter, gaseous and stellar components.

\subsubsection{Gas and stars in the haloes}\label{sect:evolution}
\begin{figure*}
\centering
\includegraphics[width=1.05\textwidth]{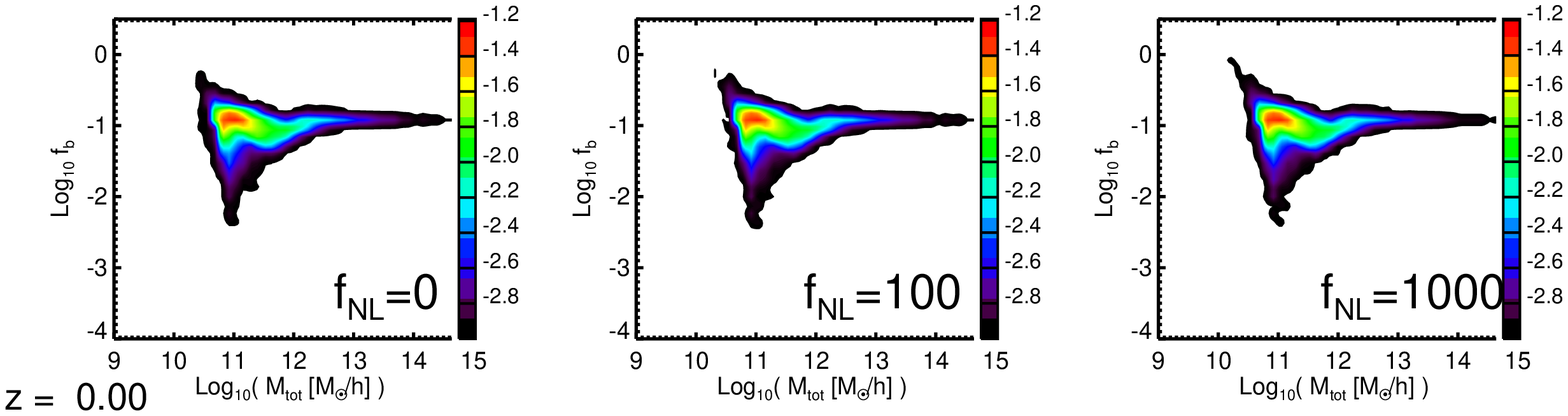}
\includegraphics[width=1.05\textwidth]{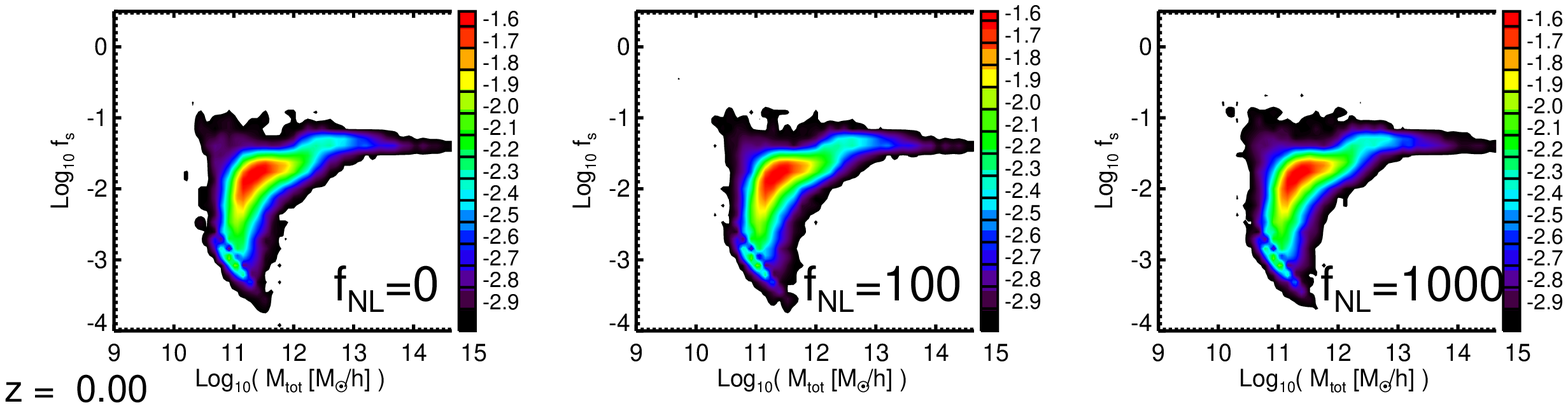}
\caption[Baryon fractions]{\small
Baryon fraction, $f_b$ (upper panels), and stellar fraction, $f_s$ (lower panels), for all the halo total masses, $\rm M_{tot}$, at redshift $z=0$, in the simulations with \fnl=0 (left), \fnl=100 (center), and \fnl=1000 (right), color-coded according to their probability distribution (in logarithmic scale), $P(f_b, M_{tot})$, and $P(f_s,M_{tot})$, respectivelly. These results refer to the top-heavy popIII IMF cases.
}
\label{fig:fb}
\end{figure*}

\begin{figure*}
\centering
\includegraphics[width=1.05\textwidth]{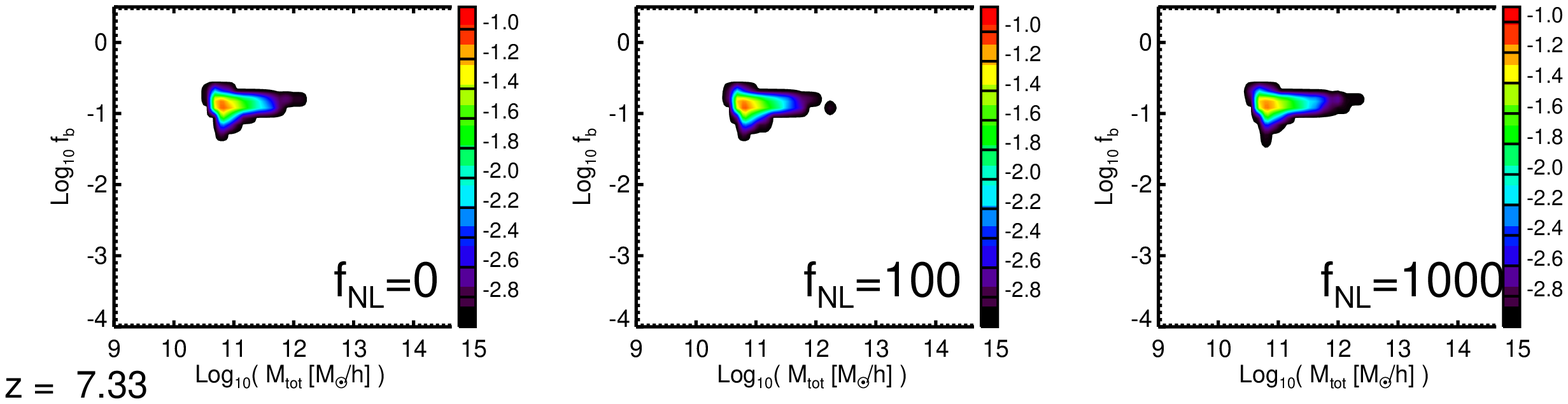}
\includegraphics[width=1.05\textwidth]{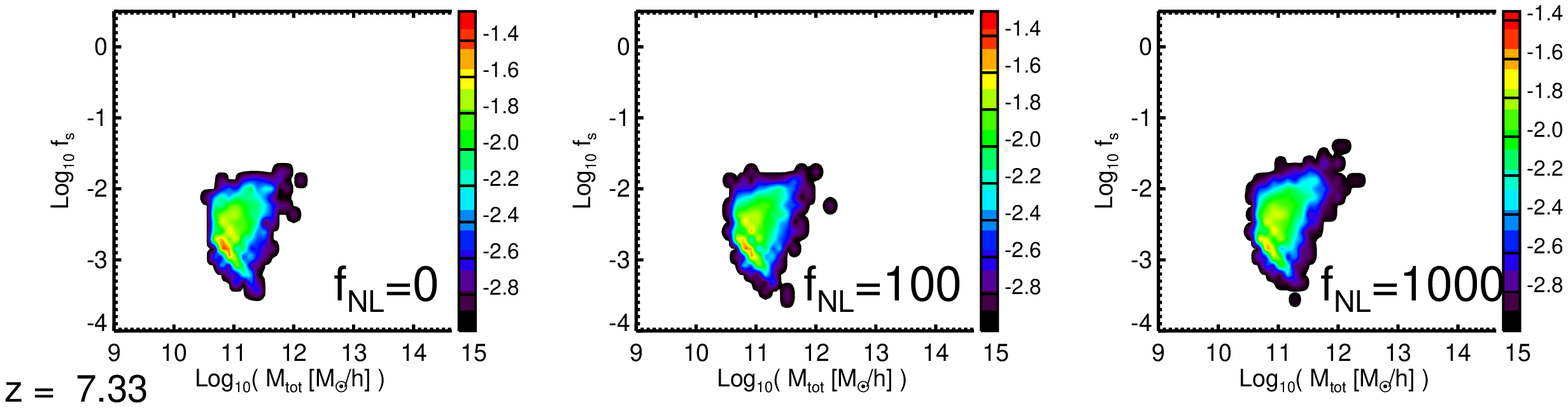}
\caption[Baryon fractions]{\small
Same as Fig.~\ref{fig:fb}, for redshift $z=7.33$.
}
\label{fig:fb2}
\end{figure*}

In Fig.~\ref{fig:fb}, the joint probability distributions $P(f_b, M_{tot})$ and $P(f_s, M_{tot})$ for the baryon fraction, $f_b$ (upper panels), and for the stellar fraction, $f_s$ (lower panels), respectivelly, as a function of the halo total mass, $M_{tot}$, at redshift $z=0$, are displayed:
\begin{equation}
\label{eq:pb_diff}
P(f_b, M_{tot}) = \frac{1}{N_{h,sample}} \frac{{\rm d} N_h}{{\rm d} M_{tot}~{\rm d} f_b},
\end{equation}
and
\begin{equation}
\label{eq:ps_diff}
P(f_b, M_{tot}) = \frac{1}{N_{h,sample}} \frac{{\rm d} N_h}{{\rm d} M_{tot}~{\rm d} f_s}.
\end{equation}
In both these relations, $N_{h,sample}$ is the total number of haloes sampled by the simulations, while $\d N_h$ is, in equation (\ref{eq:pb_diff}), the number of haloes within a given total-mass bin, $\d M_{tot}$, and within a given baryon-fraction bin, $\d f_b$, and, in equation (\ref{eq:ps_diff}), the number of haloes within a given total-mass bin, $\d M_{tot}$, and within a given stellar-fraction bin, $\d f_s$.
With the previous definitions, the distributions are correctly normalized to 1.
\\
For the three \fnl{} cases, we consider the runs with top-heavy popIII IMF.
\\
In Fig.~\ref{fig:fb}, both baryon and stellar fractions have similar behaviours and no big differences are found.
The baryon fraction, $f_b$, almost approaches the cosmic value at larger masses (a more quantitative analysis is performed in the next section) and exhibits some scatter (because of feedback mechanisms) in smaller structures, with peak values around $f_b\sim 10^{-1}$.
The stellar fraction, $f_s$, has a much larger scatter around the peak values of $f_s\sim 10^{-2}$, mostly in haloes with masses lower than $\sim 10^{12}\rm\msunh$, and achieves asymptotic values of $\sim 10^{-1.5}$ in the larger objects.
In all these trends, though, no crucial impacts from different \fnl{} are visible.
\\
The large scatter of $f_b$ and the even larger scatter of $f_s$ on galactic scales are strictly related to feedback effects that can strongly influence low-mass structures, because of their shallower potential wells.
In fact, depending on the environment, gas heated by SN explosions in small haloes can be efficiently lost, and, on the other hand, more quiescent objects can be enriched by the gas expelled from nearby structures.
Stellar scatter is probably connected to wind feedback, since each SPH particle undergoing star formation is eventually converted into a conlisionless star-like particle with a typical velocity of $\sim \rm 500\,km/s$.
Such velocities allow conlisionless stellar particles to escape from small-halo potentials and to end up into different haloes.
\\
As a comparison, in Fig.~\ref{fig:fb2}, we show the status at redshift $z\simeq 7.3$.
In this case too there are no significant changes among the \fnl{} cases considered.
Nevertheless, in the \fnl=1000 case there is slightly more baryonic (gaseous and stellar) mass in the larger objects, at masses of $\sim 10^{12}\rm\msunh$: this reflects the growth enhancement at high redshift in models with large \fnl{} values.
The baryon fraction, $f_b$, peaks at $\sim 10^{-1}$, quite close to the cosmic value $f_{b,cosmic} = \Omega_{\rm b}/\Omega_{\rm m}\simeq 0.1333$, for our choice of parameters.
Since the star formation history at $z\sim 7.3$ has only started from just $\sim 0.5\,\rm Gyr$, the stellar fraction peaks at about one order of magnitude below the values at $z=0$, i.e. at $f_s\sim 10^{-3}-10^{-2.5}$.
These results are almost independent from \fnl{}, also in such high-$z$ regimes, and keep some non-Gaussian signatures only at large masses.
\\
As a consequence, different \fnl{} might slightly affect the masses of the haloes (see also Sect.~\ref{sect:massivehalo}), but not significantly the gaseous and stellar fractions.
In all the cases, the spread at low masses (at less than $\sim 10^{12}-10^{13}\,\msunh$) is independent from \fnl, since it is similarly large and it becomes smaller and smaller at masses corresponding to galaxy groups or clusters, with the same trend for any \fnl.
This is essentially due to the capability of feedback in heating or expelling the gas from lower potentials, and enriching the closeby satellites.
In larger objects, the material is more easily trapped and minor losses take place.
In support of this, we will see in the next section that substantial amounts of baryons are lost during the bulk of cosmic star formation episodes, when feedback mechanisms become very important.

\subsubsection{Evolution of gaseous and stellar components}\label{sect:massivehalo}

\begin{figure*}
\centering
\includegraphics[width=0.32\textwidth]{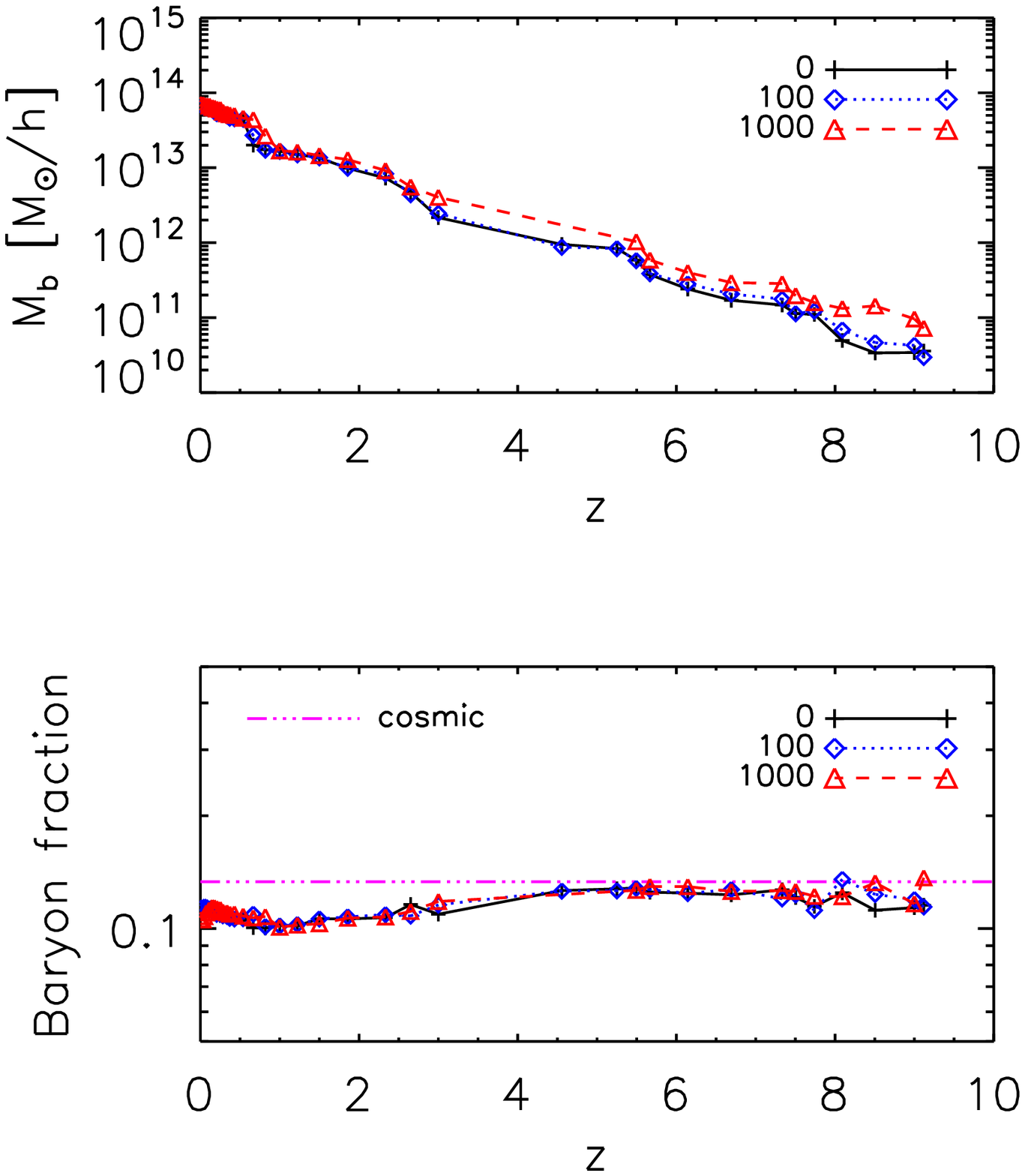}
\includegraphics[width=0.32\textwidth]{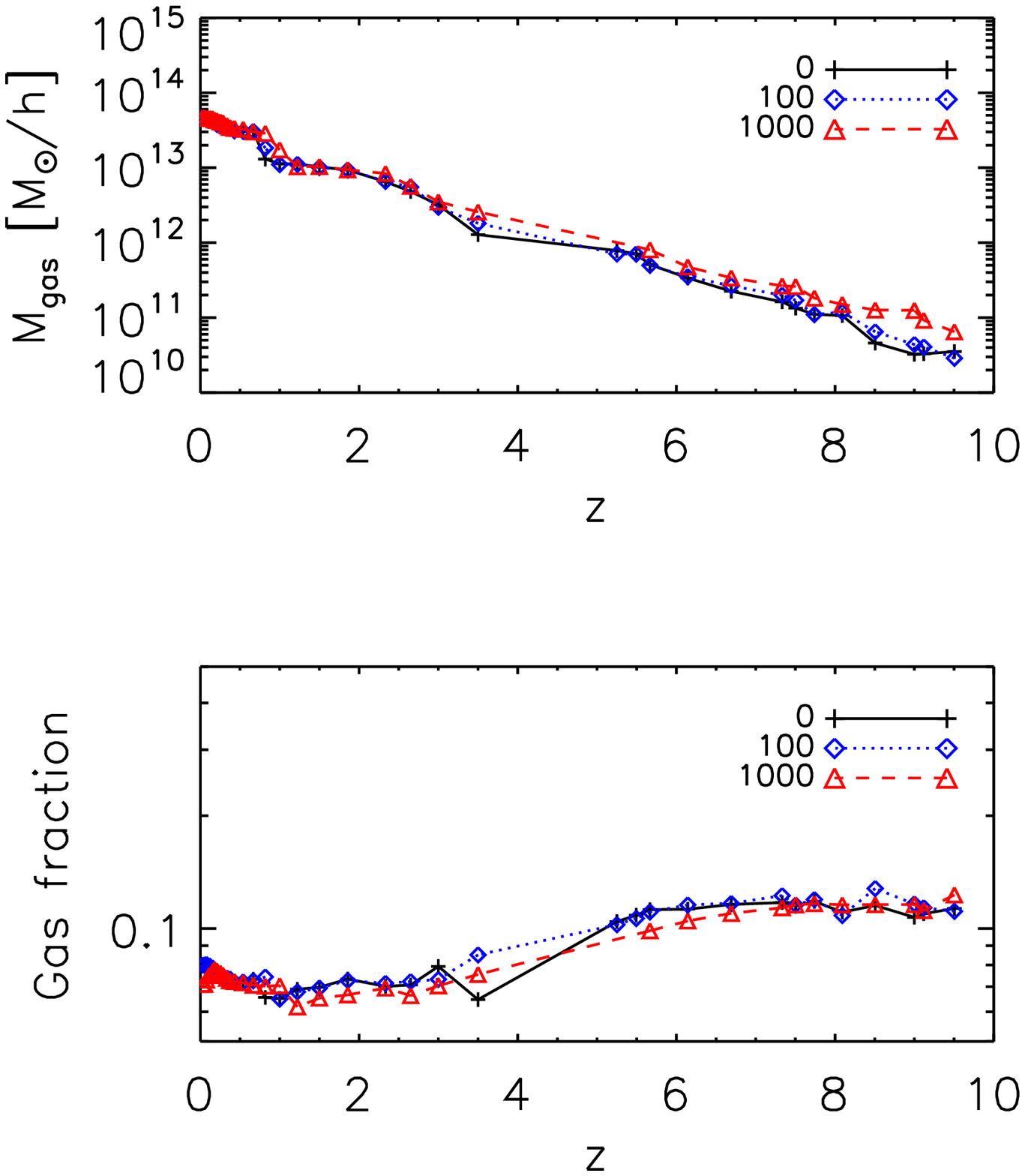}
\includegraphics[width=0.32\textwidth]{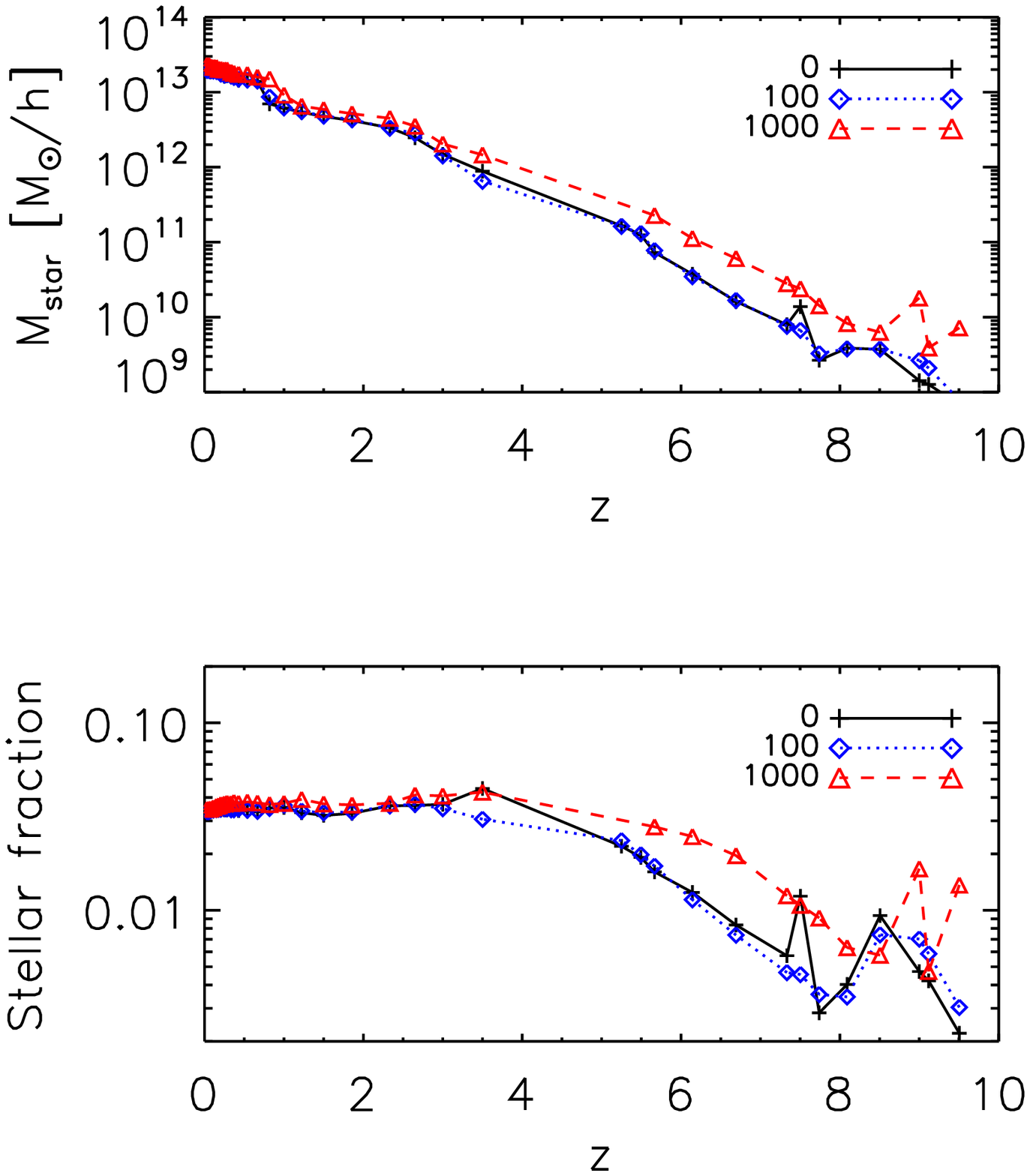}
\caption[Stellar evolution]{\small
Redshift evolution of the masses (upper) and fractions (lower) of the baryonic (left), gaseous (center), and stellar (right) components, in the most massive halo, for \fnl=0 (solid lines), \fnl=100 (dotted lines), and \fnl=1000 (dashed lines).
Results refer to the top-heavy popIII IMF cases.
The dot-dot-dot-dashed line in the bottom left panel is the cosmic baryon fraction expected from the simulations, $f_{b,cosmic} = \Omega_{\rm b}/\Omega_{\rm m}$.
}
\label{fig:baryonmass.1}
\end{figure*}

\begin{figure*}
\centering
\includegraphics[width=0.32\textwidth]{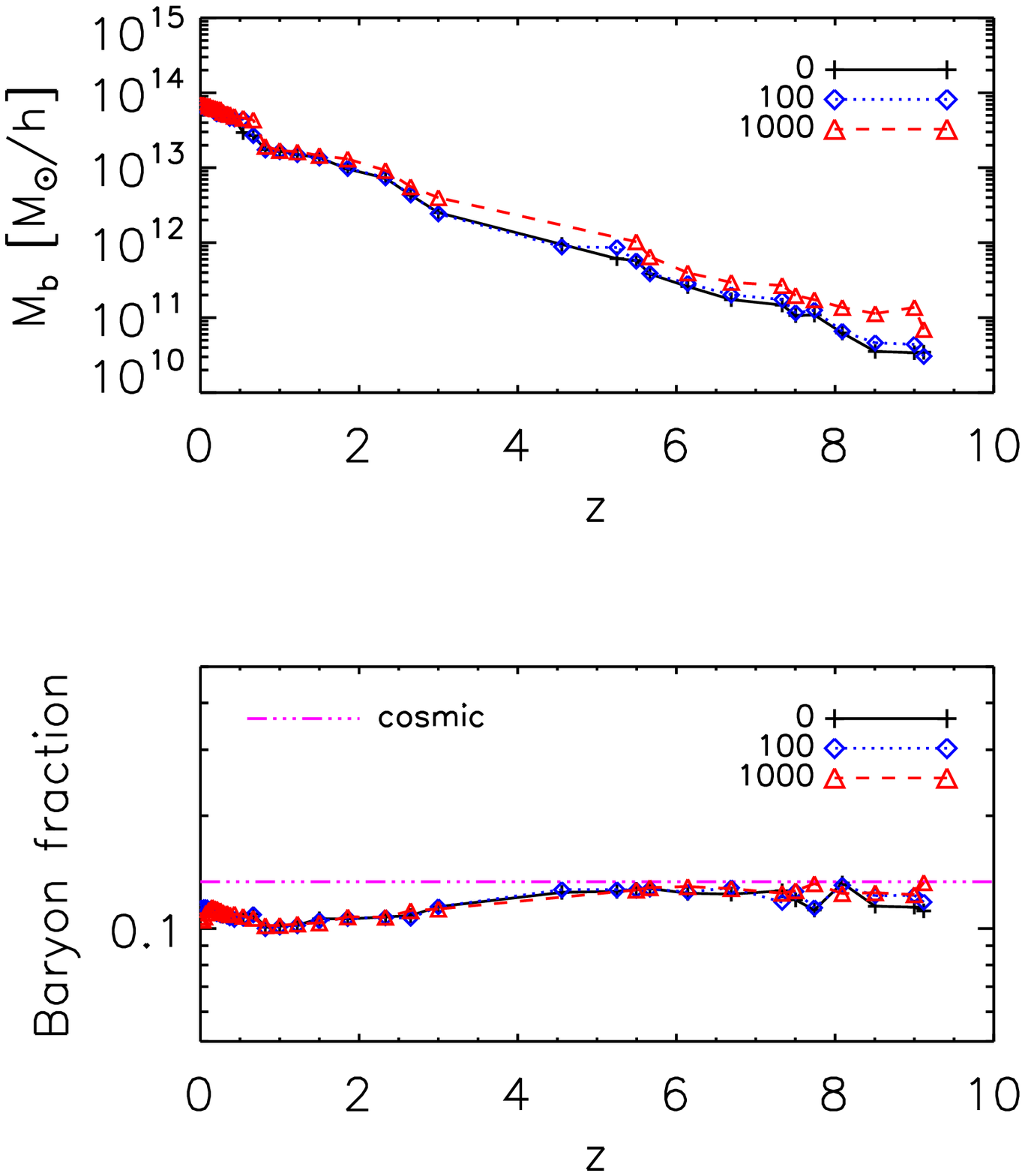}
\includegraphics[width=0.32\textwidth]{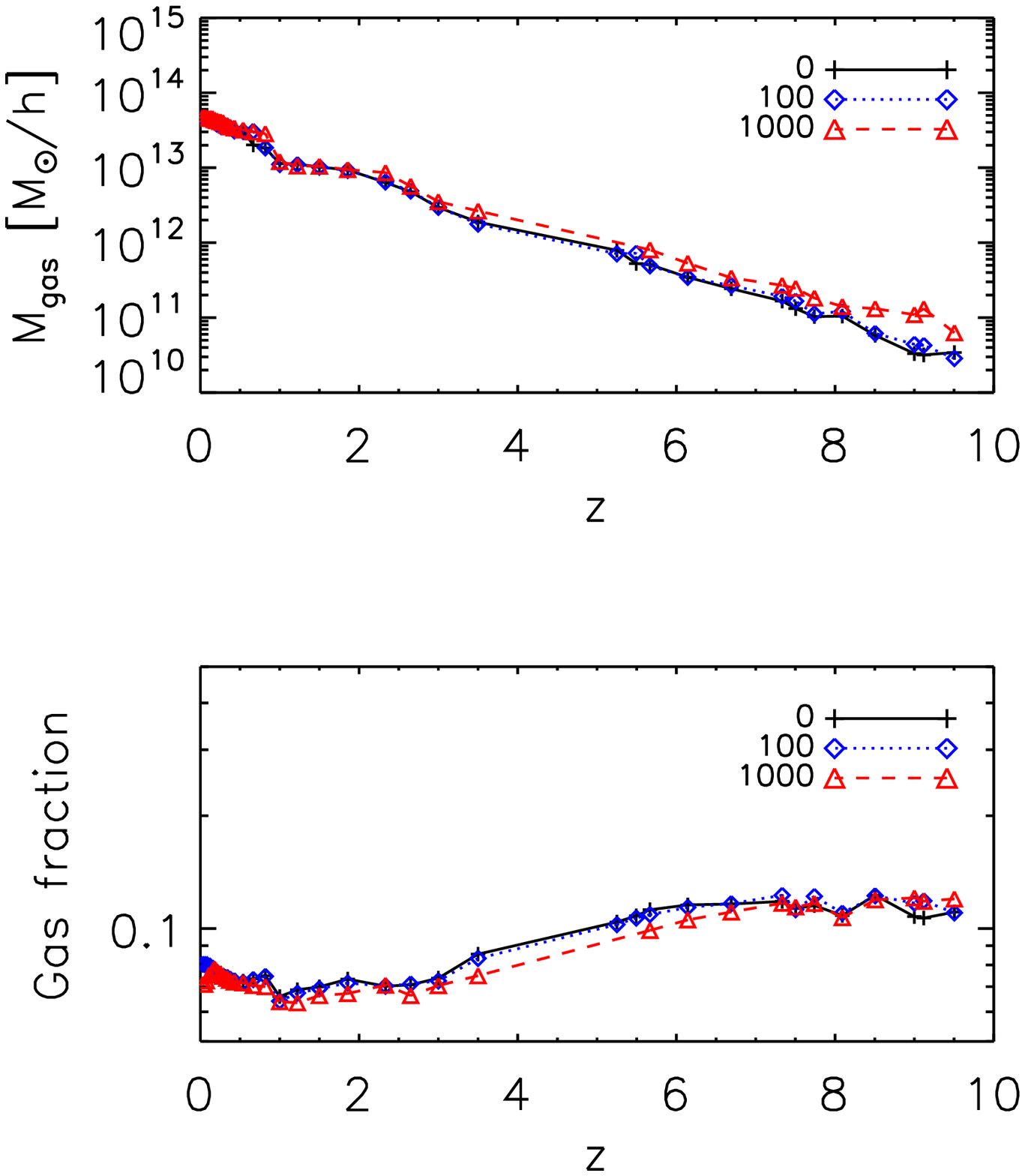}
\includegraphics[width=0.32\textwidth]{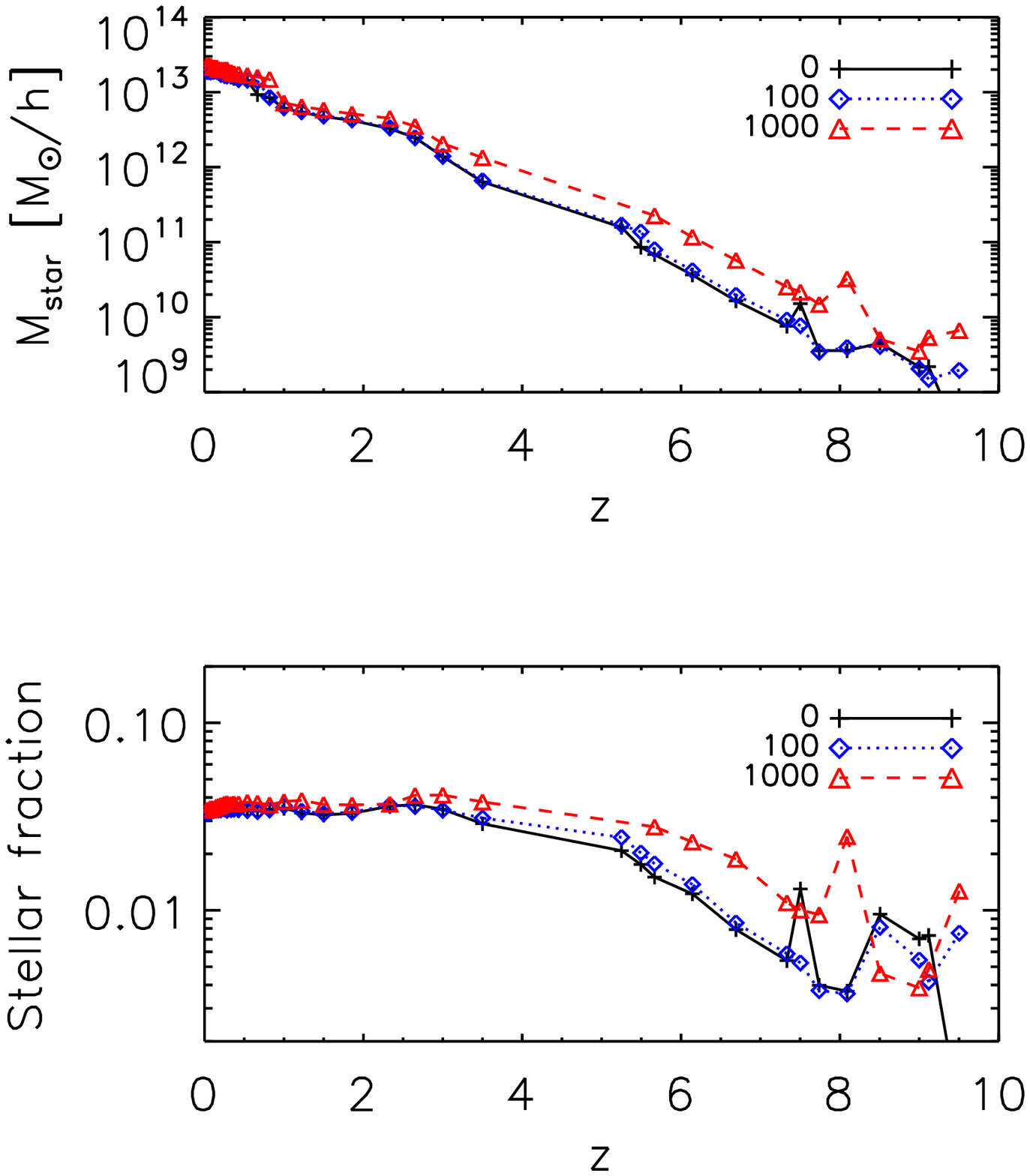}
\caption[Stellar evolution]{\small
Same as Fig.~\ref{fig:baryonmass.1}, for a Salpeter-like popIII IMF.
}
\label{fig:baryonmass.2}
\end{figure*}

To better check evolutionary differences due to the assumed value for \fnl, we follow the gaseous and stellar components within the most massive haloes found in the simulations, where also most of the star formation activity takes place.
\\
In Fig.~\ref{fig:baryonmass.1}, we plot masses (upper panels) and fractions (lower panels) of the baryonic (left), gaseous (center), and stellar (right) components, for the runs with \fnl=0, \fnl=100, \fnl=1000.
The assumed popIII IMF is a top-heavy one, while the dot-dot-dot-dashed line in the bottom left panel is the cosmic baryon fraction expected from the simulations.
The baryon mass is about a few times $10^{10}\,\msunh$ at $z\sim 10$, and $\sim 10^{14}\,\msunh$ at $z=0$, hence this fairly represents the evolutionary path of a galaxy-cluster-like object.
Also in this case, the mass evolution suggests that the model with the larger \fnl=1000 predicts larger masses and earlier growth in all the components, while the models with \fnl=0 (Gaussian) and \fnl=100 do not have significant deviations.
Looking at the temporal sequence, one sees that the major effects of non-Gaussianities are at high redshift (larger than $z\sim 3$), where differences reach up to a factor of $\sim 2-3$ ($\sim 10$ in the very early phase) for the stellar mass, as highlighted in the bottom panels.
Gas masses, instead, are affected by just $\sim 1\%-10\%$, at most.
The baryon fraction evolution shows that $f_b$ is roughly constant at $z > 5$ and traces the cosmic baryon fraction, $f_{b,cosmic}$, quite well at early times.
However, there are evident deviations at $z$ below $\sim 5$ that reach the largest discrepancy between $z\sim 1-3$.
The reason for that is immediately explained by Fig.~\ref{fig:SFR}.
Indeed, at those $z$ the star formation rate has its peak, and feedback mechanisms become more and more effective in removing material from the star forming sites, and in regulating the star formation activity (as visible e.g. from the gas and stellar fraction trends in the bottom-center and bottom-right panels).\\
Thus, star formation, also triggered by merger episodes, and the related feedback mechanisms are expected to be important sources of discrepancy between the cosmic baryon fraction and the one observationally estimated in galaxy groups and clusters \cite[e.g.]{Giodini2009}.
According to our findings, gas evacuation reduces $ f_b$ of about $25\%-30\%$, mostly during the cluster assembly history, at $z > 1$, when smaller objects can loose more easily the material heated by feedback.
E.g., an object with a baryon mass of $\sim 10^{14}\,\rm \msunh$ at redshift $z=0$ (as the one in Fig.~\ref{fig:baryonmass.1}) is found to have a mass a few times smaller at redshift $\sim 1$ and more than ten times smaller at redshift $z\sim 3$, when its gas is being lost and the corresponding baryon fraction decreases.
At $z < 1$, the gas results more tightly bound to the growing potential and some additional merging material, together with the decreasing star formation rate, sustains $f_b$.
\\
The role of non-Gaussianities in these regimes has little relevance, since, due to feedback effects on hydrodynamics, the high-$z$ enhancements of $f_s$ (due to large \fnl{} values) gradually disappear and are completely absent below $z\sim 3$.
The final values are always around $f_b\sim 0.1$, and $f_s\sim 0.03-0.04$.
\\
In Fig.~\ref{fig:baryonmass.2}, a similar analysis for the Salpeter-like popIII IMF is done, but, besides the aforementioned differences at very early times ($z\sim 8-10$), the trends are still very similar and the physical interpretation does not change.

\subsubsection{Profiles}\label{sect:profiles}

\begin{figure}
\centering
\includegraphics[width=0.7\textwidth]{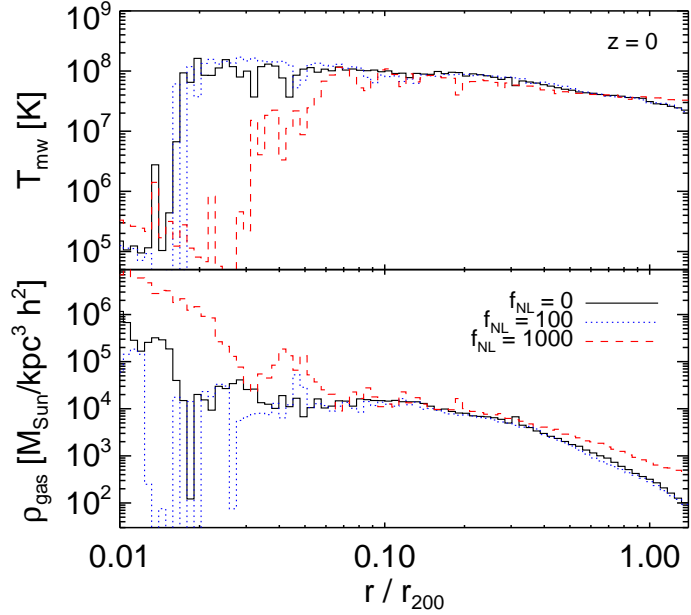}
\caption[Profiles]{\small
Mass-weighted temperature (upper) and gas density (lower) profiles for the most massive halo found in the runs with \fnl=0 (solid lines), \fnl=100 (dotted lines), and \fnl=1000 (dashed lines).
}
\label{fig:profiles}
\end{figure}

Finally, in Fig.~\ref{fig:profiles} the radial ($r$) profiles at $z=0$ for the gas mass-weighted temperature ($T_{mw}$) and density ($\rho$) of the most massive object are displayed in the three cases \fnl=0, \fnl=100, and \fnl=1000 (for a top-heavy popIII IMF).
The $x$-axis is $r/r_{200}$, with $r_{200}$ the virial radius\footnote{
The virial radius $r_{200}$ is defined as the radius at which the matter density is 200 times the critical density.
} of each halo, equal to 1.064~\Mpch, 1.131~\Mpch, 0.736~\Mpch, for \fnl=0, \fnl=100, \fnl=1000, respectively.
The gas distribution in the haloes reaches overdensities of the order of $\sim 10^4-10^5$ (the critical density at $z=0$ is about $277.4\,\rm M_\odot/kpc^3 {\it h}^2$), so the baryons are essentially decoupled from the cosmological frame.
The global behaviour in the three cases is quite similar with a low-temperature environment ($T_{mw}\sim 10^5-10^6\,\rm K$) in the high-density regions (representing the cold innermost core) and an asymptotic behaviour at larger radii (where $T_{mw}\sim 10^{7}\,\rm K$).
We notice that, apart from the denser core in the \fnl=1000 model, no other clear trends with \fnl{} are visible, and that the noisy inner part is more likely dominated by stochastic star formation feedback, gas cooling, non-thermal motions, etc. \cite{Sarazin1988,Borgani2009arXiv,Fang2009,Lau2009,Biffi2011}.
As a consequence, it seems that these structures have lost memory of the primordial initial non-Gaussianities and that their thermodynamical properties are mainly led by hydrodynamics and feedback mechanisms.
These are typical examples of how the effects of different \fnl{} values can be strongly mitigated or even cancelled by the more powerful effects coming from gaseous and stellar evolution within the growing haloes.
\\
We also point out that the drops in the temperature profiles at distances of a few per cent $r_{200}$ could be presumably alleviated by additional heating processes not explicitly considered here, like thermal conduction, cosmic rays or AGN feedback, and that are still widely debated in literature.
An extensive study of all these different phenomena is beyond the aims of this work, but regarding non-Gaussianities we can safely state that they do not represent a crucial caveat, since they would add more degrees of stochasticity and would not affect our conclusions on the implications from different \fnl{} parameters.


\section{Summary and conclusions}\label{sect:discussion}

In this work, we have considered the first large-scale N-body/hydrodynamical numerical simulations, dealing with non-Gaussian initial conditions, and including gas cooling, star formation, stellar evolution, chemical enrichment (from both population III and population II regimes), and feedback effects \cite{MaioIannuzzi2011}.
We have studied in detail the basic properties and evolution of baryons in the different scenarios, by paying attention to the effects on star formation (Sect.~\ref{sect:sfr}) on gas evolution (Sect.~\ref{sect:pdf}), on metal pollution (Sect.~\ref{sect:ff}), and on the growth down to low redshift of the cosmological structures and their tgaseous and stellar content (Sect.~\ref{sect:baryons}).
\\
The star formation (Fig.~\ref{fig:SFR}) in Gaussian (\fnl=0) and non-Gaussian (\fnl=100, \fnl=1000) models presents different behaviours, mostly at high redshift and for larger \fnl, and, consistently with that, also the probability distribution functions (Fig.~\ref{fig:pdf100}) show more overdense gas in the \fnl=1000 case and less in the \fnl=100 and \fnl=0 cases, to eventually converge at low redshift.
\\
The metal enrichment process is consequently affected (Fig.~\ref{fig:ffIMF}), since metal pollution from dying stars starts earlier for \fnl=1000 and later for \fnl=100 and \fnl=0.
Different choices for the popIII IMF have effects at early times, as metals are very sensitive to the typical lifetimes of the first stars, but do not alter the trends for different \fnl.
\\
Baryon, gas and star fractions within cosmological structures are very little affected by non-Gaussianities at low redshift (Fig.~\ref{fig:fb}), but might be quite affected at high redshift (Fig.~\ref{fig:fb2}).
In all the cases, typical values have quite a large scatter at low masses ($< 10^{12}\,\msunh$ at $z=0$) and $f_b$ achieves almost the cosmic value at larger masses.
More precisely, $f_b$ stabilizes a bit below the cosmic fraction with a weakly increasing trend for galaxy groups and clusters, while $f_s$ has a scatter of a few orders of magnitude below $\sim 10^{12}\,\msunh$ and an almost constant or slightly decreasing trend at larger masses, where $f_s \sim 10^{-1.5}$ \cite[for observational studies at $z<1$, and references therein]{Giodini2009,Alex2010}.
\\
Some clear differences due to the assumed \fnl{} are more easily evident when following individual objects.
In the case of the most massive clusters, gas mass evolution are affected by only $\sim 1\%-10\%$ at most, and stellar masses by up to a factor of $\sim 2-3$ at $z> 4$, both in the top-heavy popIII IMF scenario (Fig.~\ref{fig:baryonmass.1}) and in the Salpeter-like scenario (Fig.~\ref{fig:baryonmass.2}).
These differences survive only at high redshift, though, since, the feedback mechanisms from gas and stellar evolution dominate the baryon behaviour at later times (Fig.~\ref{fig:profiles}).
\\
The baryon fraction is significantly correlated to star formation and feedback, since we find discrepancies from the cosmic values (of $\sim 25\%-30\%$) only when star formation becomes relevant ($z<4$), independently from the primordial dark-matter distribution.
\\
Thus, it is likely that feedback mechanisms are the main responsible for washing out non-Gaussian signatures on the baryonic matter, by heating and mixing gas and by enhancing its turbulent state \cite{Borgani2008}.
This would also mean that other gas properties, like temperature, density, velocity, can not keep vivid memory of the primordial \fnl{} at lower redshift.
The 21-cm emission from high-$z$ galaxies might still preserve some information, during the first Gyr, but the expected impact on, e.g., the X-ray emission from galaxy clusters or on the signal from the SZ effect(s) \cite{KitayamaSuto1997,Borgani1999,Sartoris2010} should be as small as the ones on the baryons, since they involve quantities, like electron fraction and temperture, that are strongly dominated by the feedback processes.
\\
As already mentioned in the previous sections, what emerges from our analysis is the need for high-redshift research, in order to investigate purely non-Gaussian effects.
Moreover, thanks to available high-energy satellites (as e.g.
$Swift$\footnote{http://heasarc.nasa.gov/docs/swift/swiftsc.html}, 
$Fermi$\footnote{http://fermi.gsfc.nasa.gov/}, 
$Integral$\footnote{http://science1.nasa.gov/missions/integral/}), it is becoming possible to explore the $z\sim 10$ Universe also observationally \cite{Cucchiara2011arXiv}, and to impose constraints (based on observed data) on the GRB rate at high $z$ \cite{Campisi2011}.
Since the latter is strictly linked to the star formation history of the Universe and to its underlying cosmological model, GRBs might be suitable cosmological probes of non-Gaussianities, independent from cosmic microwave background (CMB) determinations.
Also high-$z$ galaxies and quasars, could represent interesting probes of non-Gaussianities.
The former are observed up to $z\sim 10$, and the latter are detected up to $z > 6$ together with many young, star-forming galaxies \cite{Fan2001,YanWindhorst2004,Cool2006,Bouwens2007,Utsumi2010}.
Since these early structures are supposed to be tracers of high-density regions at high $z$ \cite{Djorgovski2003,Overzier2009}, it is likely that their stellar content could be quite sensitive to non-Gaussianities, too.
\\
We conclude our discussion by summarizing as follows:\\
(i) non-Gaussianities lead to an earlier evolution of primordial gas, structures, and star formation;\\
(ii) the consequent metal enrichment starts earlier (with respect to the Gaussian scenario) in non-Gaussian models with larger \fnl;\\
(iii) gas fractions within the haloes are not significantly affected by the different values of \fnl, with deviations of $\sim 1\%-10\%$;\\
(iv) the stellar fraction is quite sensitive to non-Gaussianities at early times, with discrepancies reaching up to a factor of $\sim 10$ at very high $z$, and rapidly converging at low $z$;\\
(v) the trends at low redshift are independent from \fnl{}: they are mostly led by the ongoing baryonic evolution and by the feedback mechanisms, which determine a discrepancy in the baryon fraction of galaxy groups/clusters of $\sim 25\%-30\%$ with respect to the cosmic values;\\
(vi) the impacts on the cluster X-ray emission or on the SZ effect(s) are expected to be not very large and dominated by feedback mechanisms, whereas 21-cm from high-redshift galaxies might preserve more information related to \fnl;\\
(vii) in order to explore non-Gaussianities and baryonic structure formation, high-redshift ($z\sim 10$) investigations are required, with first stars, galaxies, quasars, and GRBs being optimal fields of study, and potential cosmological probes of non-Gaussianites.


\section*{Acknowledgments}
The analysis and the simulations have been performed on the IBM Power 6 machine of the Garching Computing Center (Rechenzentrum Garching, RZG).
The simulations were run by U.~M. with initial conditions kindly provided by Francesca Iannuzzi.
We aknowledge useful discussions with D.~Pierini.
This research has made use of NASA's Astrophysics Data System.


\section*{References}

\bibliographystyle{jphysicsB}
\bibliography{bibl.bib}


\end{document}